\pdfoutput=1

\documentclass[aps,prb,reprint,showpacs,superscriptaddress]{revtex4-1}

\usepackage[T1]{fontenc}
\usepackage{graphicx}
\usepackage{amsmath}
\usepackage{amssymb}
\usepackage{dcolumn}
\usepackage{dsfont}
\usepackage{latexsym}
\usepackage{rotating}
\usepackage{color}
\usepackage{latexsym}
\usepackage{bbm}
\usepackage{subfigure}
\usepackage{float}
\usepackage{epsfig}
\usepackage{psfrag}
\usepackage{natbib}
\usepackage{bm}
\usepackage{amsthm}
\usepackage{eucal}
\usepackage{mathrsfs}
\usepackage{url}

\usepackage{color} 

\usepackage{hyperref}
\hypersetup{
colorlinks=true,final=true,
        linkcolor=red,
        citecolor=blue,
        filecolor=blue,
        urlcolor=blue,
}

\begin{document}


\title{Emergent Spin Hall phase at a Lifshitz transition from competing orders}

\author{N. Mohanta}
\email{nmohanta@phy.iitkgp.ernet.in}
\affiliation{Department of Physics, Indian Institute of 
Technology Kharagpur, W.B. 721302, India}
\author{S. Bandopadhyay}
\affiliation{Department of Physical Sciences, Indian 
Institute of Science Education and Research Kolkata, W.B. 741252, India}
\author{S. Lal~*}
\email{slal@iiserkol.ac.in}
\affiliation{Department of Physical Sciences, Indian 
Institute of Science Education and Research Kolkata, W.B. 741252, India}
\author{A. Taraphder}
\affiliation{Department of Physics, Indian Institute of 
Technology Kharagpur, W.B. 721302, India}
\affiliation{Centre for Theoretical Studies, Indian 
Institute of Technology Kharagpur, W.B. 721302, India}


\maketitle

{\bf{The effects  of competing  orders, such as  superconductivity and
    ferromagnetism,  on  a  Fermi  liquid  are  well  established.   A
    comprehensive understanding of such a competition in a metal whose
    Fermi   surface  has  a   non-trivial  topology   is  yet   to  be
    achieved. Here, we address this question in a prototypical system:
    the  2D  Rashba  semimetal~\cite{PhysRevLett.92.126603}.  We  show
    that dominant superconductivity  interplays with Rashba spin orbit
    interactions  (SOI) in  forming a  novel intrinsic  anomalous Hall
    effect  (AHE) with  gapless  edge states  of Bogoliubov-de  Gennes
    (BdG)   quasiparticles.     As   in   the    case   of   itinerant
    ferromagnets~\cite{RevModPhys.82.1539},  the intrinsic  AHE arises
    from  Berry  curvature  effects   in  the  band  structure.   This
    phenomenon   is  robust   even   as  sub-dominant   ferromagnetism
    dramatically      changes      the      nature     of      pairing
    symmetry~\cite{PhysRevB.77.220501,   PhysRevLett.101.160401}.   An
    emergent  spin  Hall phase  involving  a  change in  Fermi-surface
    topology is found to  accompany this Lifshitz quantum phase transition.  
    We demonstrate the coexistence  of the original and novel  AHE in the
    presence of weak  disorder.  We offer a comparison  of our results
    with  experiments on  the two  dimensional electron  gas  at oxide
    hetero-interfaces~\cite{Richter2013, Caviglia2008} as well as make
    some testable predictions.}}

AHE     has      been     observed     in      wide     variety     of
materials~\cite{RevModPhys.82.1539}    such     as    complex    oxide
ferromagnets~\cite{PhysRevB.53.4393, PhysRevB.70.180407, JPSJ.66.3893,
  PhysRevLett.93.016602},                                 ferromagnetic
semiconductors~\cite{PhysRevLett.88.207208},
spinels~\cite{JPSJ.70.2999,            Lee12032004},           Heusler
alloys~\cite{PhysRevB.66.174429,      PhysRevB.70.205114},     layered
dichalcogenides~\cite{PhysRevB.77.014433}    etc.     The    intrinsic
mechanism for  the AHE originates due to  spin-orbit interaction (SOI)
in parity-broken itinerant ferromagnets  and is understood in terms of
the  Karplus-Luttinger  semi-classical  theory~\cite{PhysRev.95.1154}.
More recently, the  intrinsic AHE has been explained  as a topological
mechanism: electrons  at the Fermi  surface can acquire a  Berry phase
from  the existence of  magnetic monopoles  in momentum  space arising
from  a non-trivial  topology of  electronic bands~\cite{Fang03102003,
  OnodaJPSJ.71.19,   PhysRevLett.88.207208}.    There  are   extrinsic
contributions  to  the  anomalous  Hall conductivity  (AHC),  such  as
side-jump~\cite{PhysRevB.2.4559}      or      skew-scattering     from
impurities~\cite{Smit1955877},  which   sometimes  dominate  over  the
intrinsic process~\cite{PhysRevLett.97.126602}.

\begin{figure*}[!ht]
\begin{center}
\includegraphics[width=170mm]{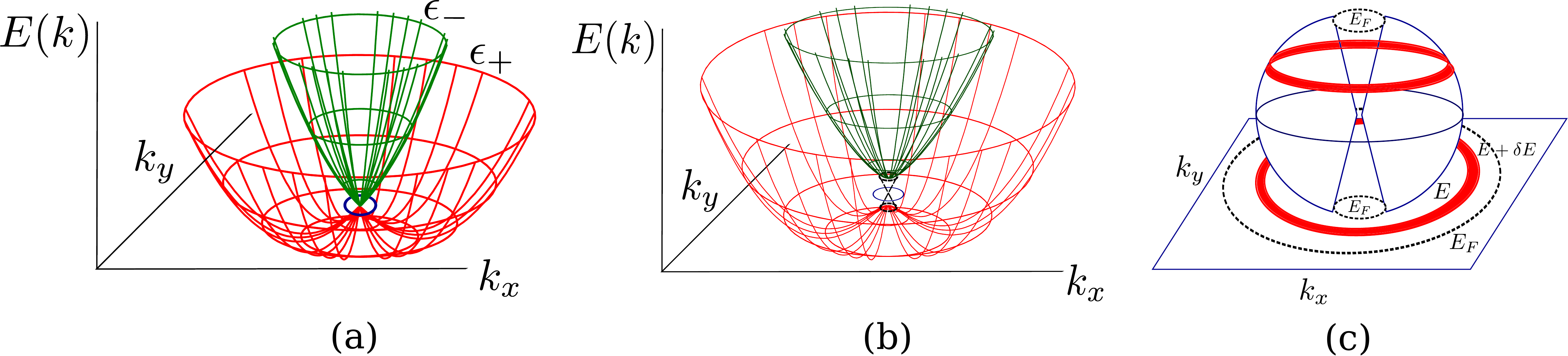}
\caption{(Color  online) {\bf The  band structure  for  the 2D  Rashba
  semimetal.} (a) A cusp-like feature connects the two bands $\epsilon_{+}$
  and   $\epsilon_{-}$,   with   a   Dirac-like  dispersion   of   the
  electrons. The  blue circle displays a  non-contractible loop around
  the Dirac  point, with a  non-trivial topology of the  Fermi surface
  characterized  by   a  Berry  phase  $\gamma=\pi$   (for  the  lower
  band). (b) The  opening of a gap leads to a  mass for the Dirac-like
  electronic  excitations.   The  gap   can  arise  due  to  a  finite
  superconducting order parameter  or magnetization. The black, dashed
  Dirac  cone  connecting the  upper  and  lower  bands signifies  the
  gapless edge-states present in any finite system.  These edge states
  are again surrounded  by a non-contractible loop leading  to a Berry
  phase $\gamma < \pi$, and make the system a Anomalous Hall insulator
  with topological properties.  (c) A stereographic projection of the
  energy contours on the  $(k_{x},k_{y})$ plane onto a sphere touching
  the plane  at its  south pole.  The  black dashed line  projects the
  Fermi energy onto a latitude  near the North pole, while states with
  energies  in  the  interval  $E$  to $E+\delta  E$  (red  band)  are
  projected  onto   the  northern  hemisphere  as   shown.   A  finite
  superconducting order parameter connects the upper boundary with the
  south pole, making it topologically equivalent to a ring torus.  }
\label{band_topol}\vspace{-2em}
\end{center}
\end{figure*}

A   particularly  simple   model  for   the  intrinsic   AHE   is  the
two-dimensional                  ferromagnetic                  Rashba
model~\cite{0022-3719-17-33-015}. Here  also, the AHE appears  due to a
Berry phase picked  up at the avoided band-crossing  induced by Rashba
SOI (see Fig.(\ref{band_topol}(a) and (b))).  The AHC  $\sigma_{xy}$ 
takes a finite value  and is proportional
to the Berry  phase (in units of  $e^2/h$) when there is a  gap at the
Fermi       level       at       the       avoided       band-crossing
point~\cite{PhysRevB.71.224423}.    However,  an   additional  singlet
superconducting  pairing gap  at  the Fermi  level  will suppress  the
magnetization induced  gap responsible  for stabilizing the  AHE. This
leads us  to expect a  clear suppression of  the intrinsic AHE  by the
singlet superconductivity.

In this  paper, we  study the intrinsic  AHE of BdG  quasiparticles in
two-dimensional   $s$-wave  superconductors   with   Rashba  SOI   and
ferromagnetism based  on a self-consistent mean-field  solution of the
pairing gap.  The system is modeled by the following Hamiltonian
\begin{equation}
\vspace{-1em}
\begin{split}\vspace{-1em}
{\cal H}&=\sum_{k,\sigma}\epsilon_k c_{k\sigma}^\dagger c_{k\sigma}
+\alpha \sum_{k,\sigma,\sigma^{\prime}}(\mathbf{g_k} \cdot \boldsymbol{\sigma} )_{\sigma \sigma^{\prime}}c_{k\sigma}^\dagger c_{k\sigma^{\prime}}\\
&- m_z\sum_{k,\sigma,\sigma^{\prime}} \boldsymbol{\sigma}_{\sigma \sigma^{\prime}}^{z}c_{k\sigma}^\dagger c_{k\sigma^{\prime}} 
+ \sum_{k}(\Delta c_{k\uparrow}^{\dagger}c_{-k\downarrow}^{\dagger}+h.c.)
\end{split}
\label{model}
\end{equation}\vspace{-0em}
where   $\epsilon_k=-2t(\cos   k_x+\cos   k_y)-\mu$   represents   the
dispersion of electrons, $t$ the hopping parameter, $\mu$ the chemical
potential,    $m_z$   the    magnetization   perpendicular    to   the
two-dimensional  plane, $\alpha$ the strength of Rashba SOI and  $\mathbf{g_k}=(\sin k_y,  -\sin  k_x)$.\,\,
$\Delta=-<c_{k\uparrow}c_{-k\downarrow}>$   is   the   superconducting
pairing gap.

We briefly recall  that for the case of  a vanishing magnetization and
$s$-wave  superconductivity,  the  Rashba  SOI leads  to  a  cusp-like
feature at the $(k_{x}=\pm\pi, k_{y}=\pm\pi)$ points in the electronic
bands; the emergent electronic excitations around this cusp correspond
to massless  Dirac fermions~\cite{PhysRevLett.92.126603} with  a Berry
phase $\gamma=\pi$ (see  Fig.~\ref{band_topol}(a)). This gives rise to
two  edge currents  in the  system with  opposite helicities,  in turn
giving  a  universal  negative   value  of  the  intrinsic  spin  Hall
conductivity (SHC)  while the intrinsic  AHC vanishes due to  an exact
cancellation.  A  finite superconducting  order parameter will  open a
gap in the quasi-particle spectrum at the Fermi surface, leading to an
avoided  band-crossing at  the cusp.   This  gives mass  to the  Dirac
fermions, leading to a Berry  phase for the BdG quasiparticles $\gamma
<\pi$  (see  Fig.~\ref{band_topol}(b)).   It  is  expected  that  this
Berry-phase contribution to the SHC  remains finite in the presence of
a superconducting  order parameter~\cite{PhysRevLett.102.086602,gradhandannett,chungroy}.  
The intra-band   and   inter-band    pairing   amplitudes   in   the   two
Rashba-split bands
$\epsilon_{\pm}({\mathbf{k}})=\epsilon_k      \pm      \xi$,     where
$\xi=(\alpha^2|\mathbf{g_k}|^2+m_z^2)^{1/2}$      are     respectively
$\Delta_{\pm} =  (-\alpha |\Delta|/(2\xi))(\sin  k_y \pm i  \sin k_x)$
($p_x \pm  ip_y$-wave)  and $\Delta_s =  m_z |\Delta|/\xi$  ($s$-wave)
(see      supplementary      information section A).       As      shown      in
Fig.~\ref{band_topol}(c),  a stereographic  projection  of the  energy
contours on the $(k_{x},k_{y})$ plane onto a sphere show that a finite
superconducting order parameter turns the sphere into a ring torus.

\begin{figure*}[!ht]
\begin{center}
\epsfig{file=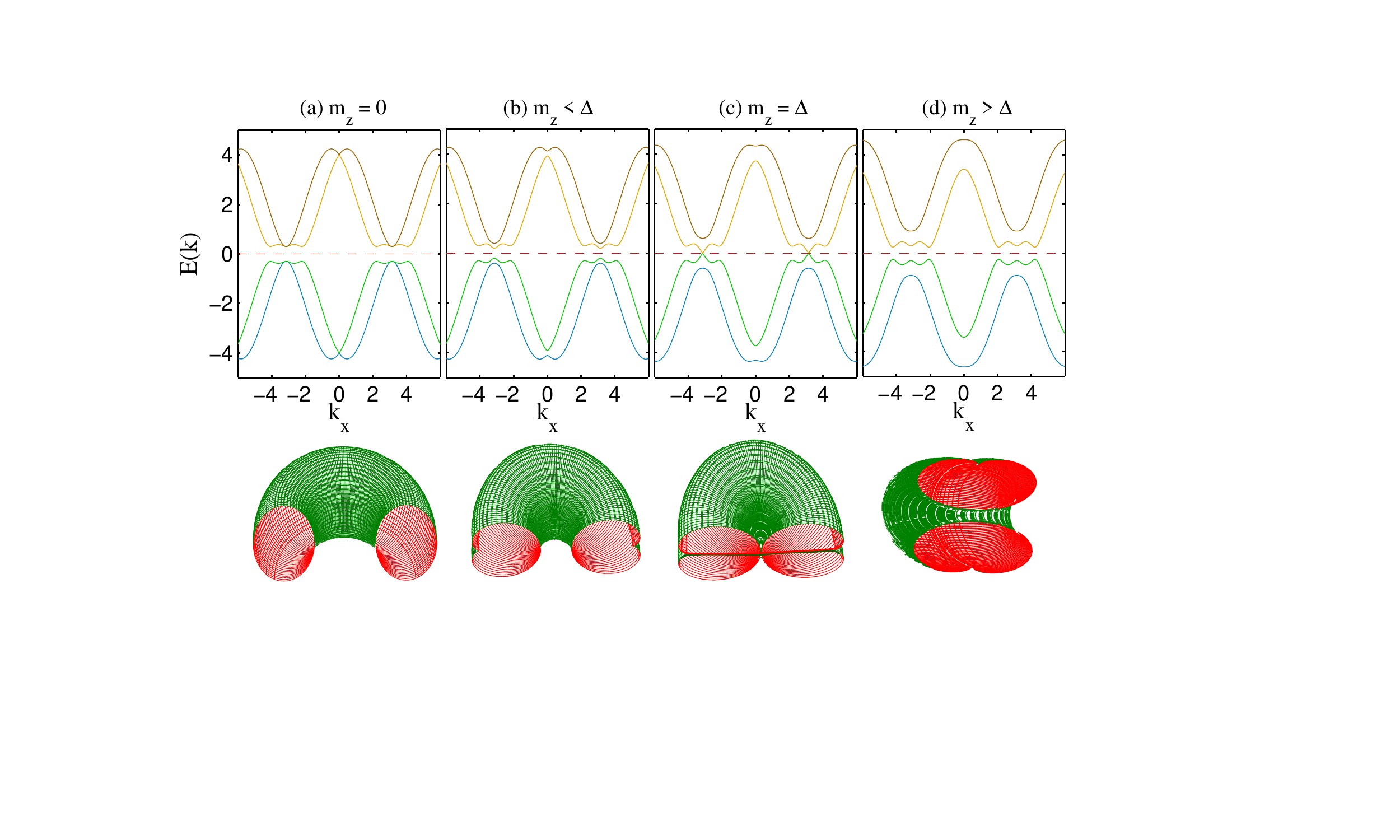, trim=1.2in 1.65in 2.3in 0.7in,clip=true, width=170mm}\vspace{-2em}
\caption{(Color online) {\bf The  Bogoliubov- de Gennes (BdG) quasi-particle spectrum.} 
  $E(k)$ as a function of $k_{x}$ and with a fixed $k_{y}=0$ for various values
  of  the  magnetization   $m_z$  with  constant  parameters  $\mu=0$,
  $\Delta=0.3$,  $\alpha=1.0$ and  $t=1$. The  avoided level-crossings
  near  $(k_{x}=\pm\pi, k_{y}=0)$ points  at the  Fermi-level give  rise  to finite
  Berry-phase $\gamma$ leading to  an AHE in the quasi-particle bands.
  Furthermore, as  $m_z$ increases, the  quasi-particle excitation gap
  reduces and becomes  zero at $m_{z}^{*}=\Delta=0.3$ at $(k_x=\pm\pi, k_{y}=0)$
  in column  (c).  With further  increase of $m_z$, a  new excitation
  gap,  proportional to  Rashba SOI,  opens  up (column  (d)) and  the
  system  undergoes   a  transition  from   topologically  trivial  to
  non-trivial  superconductor.  Figures  in lower  panel  describe the
  topology of the two BdG bands nearest the superconducting gap in the
  neighbourhood  of the $(k_{x}=\pm\pi,  k_{y}=0)$-points through
  the     stereographic     projection     method     described     in
  Fig(\ref{band_topol}(c)). (a)  As shown  in Fig(\ref{band_topol}(c)), 
  a finite superconducting order parameter turns the sphere into a
  ring torus by connecting its upper and lower boundaries. The effects
  of a  competing incipient uniform magnetization can  then be studied
  by considering  its effects  on the equatorial  plane. (b)  A finite
  magnetization gradually  closes the gap  defining the ring  torus by
  uniformly pinching  it on the  equatorial plane. For  a sub-dominant
  magnetization $m_{z}<m_{z}^{*}$, the inner hole of the ring torus is
  still   dependent   on   the   SC   order  parameter.    (c   )   At
  $m_{z}=m_{z}^{*}$,  the  inner  hole  due to  the  superconductivity
  closes and the  resulting topology is that of a  Horn Torus which is
  pinched from two orthogonal  directions. This change in the topology
  of the  torus signals the quantum phase transition discussed above, 
  coinciding with the  emergent spin Hall  phase. (d) For 
  $m_{z}>m_{z}^{*}$, the topology is once again that of a ring torus, 
  but whose inner hole is now dependent on the magnetization.}
\label{bdg_bands}\vspace{-2em}
\end{center}
\end{figure*}

Any finite magnetization in the system causes an imbalance in the spin
populations  of  the  BdG   quasiparticles  by  introducing  a  Zeeman
splitting  between the helicity  bands.  Increasing  the magnetization
from zero also leads to a gradual closing of the avoided band-crossing
at  the $(k_{x}=0,  k_{y}=\pm\pi)$ and $(k_{x}=\pm\pi,  k_{y}=0)$ 
points near  the  Fermi level.
This gives, in turn, an increase in the Berry curvature and thence the
AHC  from  the  neighbourhood  of these four points in $k$-space. 
This can  be seen by writing the  Hamiltonian (\ref{model}) in
the  Nambu-spin  basis  $\Psi=[\psi_{k}, \psi_{-k}^{\dagger}]$,  where
$\psi_{k}=[c_{k\uparrow}, c_{k\downarrow}]$, as
\begin{equation}
H(\mathbf{k})= \begin{pmatrix} \begin{array}{cc} \epsilon_k-m_{z} \sigma_z +\alpha \mathbf{g_k} \cdot \boldsymbol{\sigma} & i\Delta \sigma_y\\-i\Delta \sigma_y & -\epsilon_k+m_{z} \sigma_z +\alpha \mathbf{g_k} \cdot \boldsymbol{\sigma}^* \end{array} \end{pmatrix}
\end{equation}
From this, we obtain the quasi-particle spectrum
\begin{equation}
E(\mathbf{k})_{\pm}= \pm (\epsilon_k^2+\xi^2+\Delta^2 \pm 2\sqrt{\Delta^2 m_z^2 +\epsilon_k^2\xi^2})^{1/2}~,
\end{equation}
The quasi-particle  spectrum for various values of the magnetization $m_z$
is shown in  Fig.~\ref{bdg_bands}. 

\begin{figure}[!ht]
\begin{center}
\epsfig{file=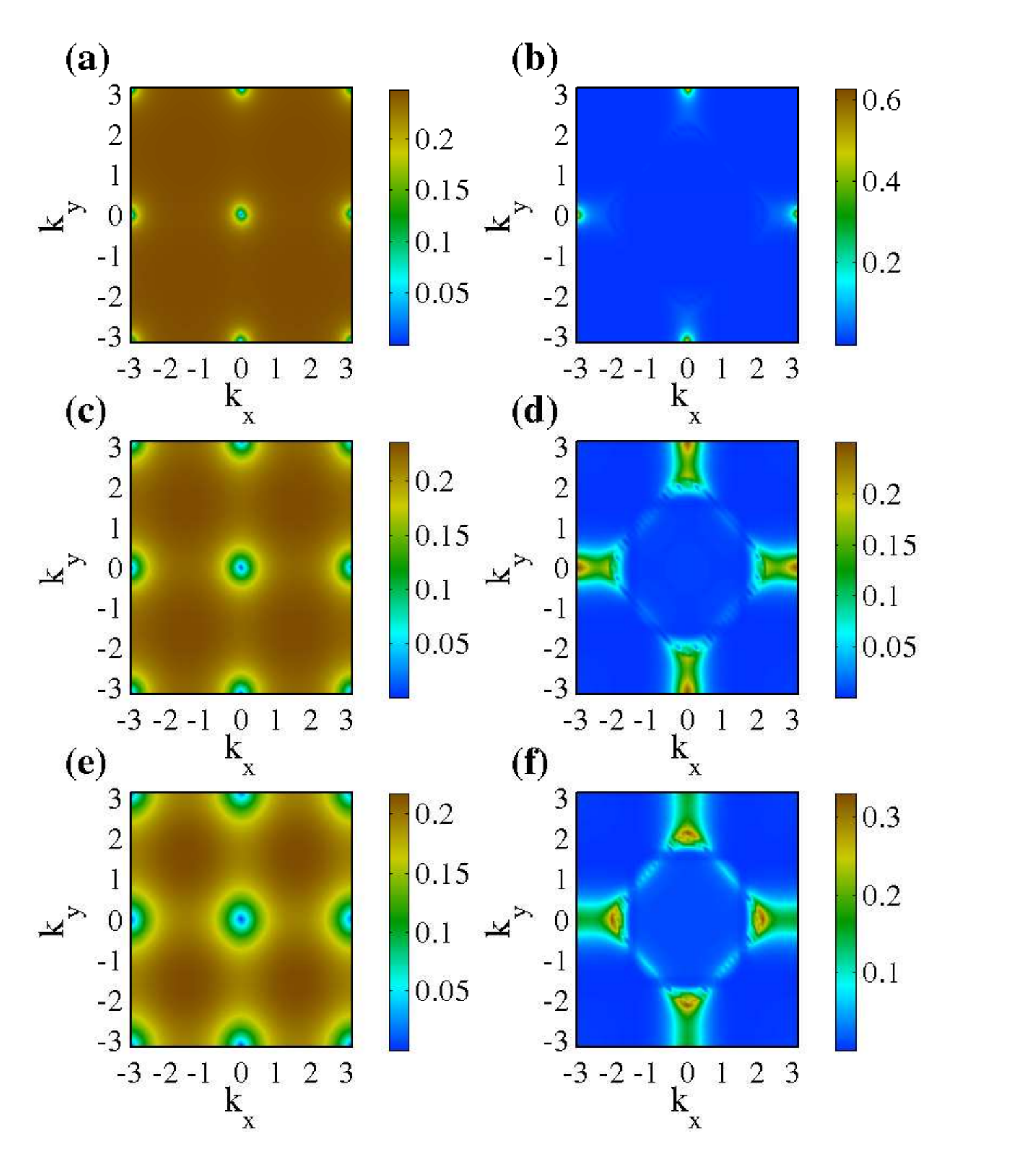, width=90mm}\vspace{-2em}
\caption{(Color  online) {\bf Momentum-space  plots of  $|\Delta_{+}|$ and $\Omega$.} The superconducting
  pairing amplitude $|\Delta_+|$ (left column) and the Berry curvature
  $\Omega$  (right  column) for  $m_z=0.2$  in  (a)-(b), $m_z=0.5$  in
  (c)-(d)  and  $m_z=0.8$  in  (e)-(f).   Other  parameters:  $\mu=0$,
  $\Delta=0.5$,  $\alpha=1.0$ and  $t=1$. $\Delta_{+}$  has  the nodal
  structure as the Rashba SOI. $\Omega$ takes finite 
  values only at the  avoided level-crossing  at  Fermi level  in the  
  quasi-particle spectrum. This is observed, for instance, at 
  $(k_{x}=0, k_{y}=\pm\pi)$ and $(k_{x}=\pm\pi, k_{y}=0)$ in (d).}
\label{kmap}\vspace{-2em}
\end{center}
\end{figure}

When the  chemical potential  $\mu$ is inside  the gap induced  by the
avoided band-crossing,  the Berry  curvature $\Omega$ and  Berry phase 
$\gamma$  picked  up  due  to  the lower  band  (with $p_x+ip_y$-wave pairing) 
is  given  by (see supplementary information section B for details)
\begin{eqnarray}
\Omega &=& \frac{m_{z}^{2}\alpha^{2}|\Delta|^2\epsilon_{+}\cos k_{x} \cos k_{y}}{8\xi^{4}(\epsilon_{+}^{2}+\Delta_{+}\Delta_{-})^{3/2}}~,~
\gamma = \frac{1}{2\pi}\iint_{B.Z.}\hspace*{-1.7em}d^{2}\vec{k}~\Omega (\vec{k})~, 
\end{eqnarray}
The             AHC             is             obtained             as
$\sigma_{xy}=\frac{e^2}{2h}\gamma$~\cite{PhysRevB.68.045327,PhysRevB.71.224423}.
The Berry phase  contribution from the other BdG  bands (whose pairing
is $p_x-ip_y$-wave) is suppressed  by the magnetization induced energy
separation   between  the   BdG  bands   of  opposite   helicity.   In
Fig.~\ref{kmap},  the  pairing amplitude  $\Delta_{+}$  and the  Berry
curvature $\Omega$ are shown in the first Brillouin zone (BZ) for various magnetization
$m_z$. $\Delta_{+}$ has  nodal structure due to the  Rashba SOI, while
$\Omega$ shows finite value near  the Fermi level. Further, the change
in  the topology  of the  band structure  in the  neighbourhood  of the
$(k_{x}=\pm\pi,  k_{y}=0)$ points  is shown in the  lower panel of
Fig.~\ref{bdg_bands}: the  effects of the  competition between gaps
induced by $\Delta$ and $m_{z}$ are seen to change the topology of the
BdG bands nearest to the superconducting gap via a Lifshitz transition~\cite{lifshitz}. 
A similar change happens at the $(k_{x}=0, k_{y}=\pm\pi)$ points as well.

\begin{figure}[!ht]
\begin{center}
\epsfig{file=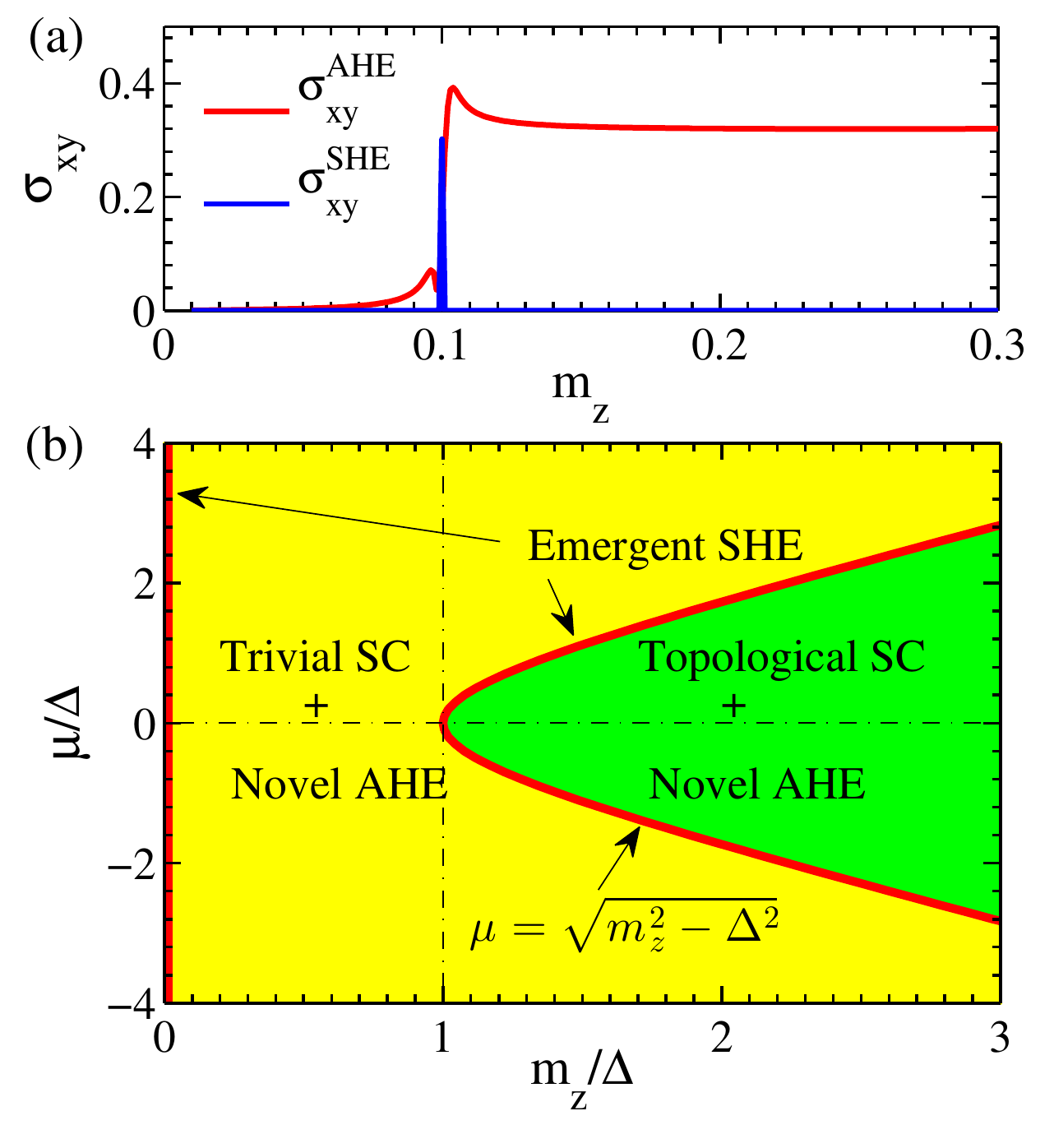, width=80mm}\vspace{-2em}
\caption{(Color  online)  {\bf Hall conductances and Phase diagram.} 
  (a)  The  variation of  the  anomalous  Hall
  conductivity  $\sigma_{xy}^{AHE}$  and  the  spin-Hall  conductivity
  $\sigma_{xy}^{SHE}$  with magnetization  $m_z$ for $\Delta=0.1$ and 
  $\mu=0$. $\sigma_{xy}^{SHE}$
  shows  a delta-function like  peak only  at $m_z=m_z^*$.   (b) Phase
  diagram  showing the  appearance of  different Hall  phases  and the
  superconductivity. It  is evident that the emergent  SHE will appear
  only  at   $m_z=0$  and  along   the  critical  points   defined  by
  $\mu=\sqrt{{m_z}^2-\Delta^2}$   line.  On  both   the  topologically
  trivial   (yellow    region)   and   non-trivial    (green   region)
  superconducting phases, the novel AHE is present.}
\label{ahe_phase}\vspace{-2em}
\end{center}
\end{figure}

We stress  that the AHE  under consideration here is different  from that
obtained in the ferromagnetic Rashba model~\cite{0022-3719-17-33-015}:
the gap at the Fermi level in the latter is opened by Zeeman splitting
whereas  in the  former,  the  pairing gap  serves  the same  purpose.
Precisely  at   a  critical  value  of  the   magnetization  given  by
$m^{*}_{z}=\sqrt{\Delta^{2}+\mu^{2}}$,    where   $\Delta$    is   the
superconducting order  parameter and $\mu$ the  chemical potential, we
find massless  Dirac fermions at the $(k_{x}=\pm\pi,  k_{y}=0)$ 
and $(k_{x}=0,  k_{y}=\pm\pi)$
points associated  with an emergent SHE  of the BdG quasiparticles. 
This can  be seen in  Fig.~\ref{bdg_bands} with 
$m^{*}_{z} (\mu=0)=\Delta$,  as well as from  the effective low-energy
Hamiltonian in the neighbourhood of $m_{z}^{*}$ (see supplementary
  information section C for details)
\begin{equation}
H_+(\mathbf{k})=(v\sin{k_y})\sigma_x+(v\sin{k_x})\sigma_y+(m_z^*-m_z)\sigma_z
\label{effcritHam}
\end{equation}
where the effective velocity of the emergent Dirac quasiparticles is 
given by $v=\alpha~(1-\frac{\mu^{2}}{m_{z}^{*2}})^{1/2}$. The mass 
of these quasiparticles is clearly seen to vanish at $m_{z}=m_{z}^{*}$. 
It is important to note that this effective low-energy subspace is 
formed out of two admixtures involving all four quasiparticle 
bands. The anomalous Hall and spin Hall conductivities computed for 
this system with a fixed $\Delta=0.1$, $\mu=0$ and with varying $m_{z}$ 
is shown in Fig.~\ref{ahe_phase}(a). A finite SHC appearing sharply at 
the critical magnetisation $m_{z}^{*}$ coincides with a vanishing AHC. 
This discontinuous behaviour of the AHC is a signature of a phase 
transition. The small value of the AHC for $m_{z}<m_{z}^{*}$ arises 
from the dominance of the $s$-wave superconductivity over ferromagnetism, 
while the sharp rise of the AHC in the regime $m_{z}>m_{z}^{*}$ signals 
the advent of a new SC order parameter discussed below. 
As shown in the mean-field phase diagram Fig.~\ref{ahe_phase}(b), the 
emergent  SHE is concomitant
with  a  quantum phase  transition  from  normal superconductivity  to
topological superconductivity  due to  the dramatic change  in pairing
symmetry  in  the  presence  of  Rashba  SOI~\cite{PhysRevB.77.220501,
  PhysRevLett.101.160401}.   This is understood  by noting  that, upon
tuning the magnetization  in the regime $0< m_{z}<  m_{z}^{*}$, one of
the helicity-bands  is pushed away from the  superconducting gap edge,
and the  inter-band $s$-wave pairing  is thus strongly  suppressed.  For
$m_{z}>m_{z}^{*}$, this  leads the system into  a phase taken  to be a
canonical example of  topological superconductivity: a superconducting
state with spinless $p_x+ip_y$-wave pairing belonging to the effective
low-energy Hamiltonian
\begin{equation}
H_+(\mathbf{k}) = \begin{pmatrix} \begin{array}{cc} \epsilon_+ & \Delta_{+} \\ \Delta_{+}^* & -\epsilon_+ \end{array} \end{pmatrix}~.
\label{H_low}
\end{equation}
Thus, beyond $m_{z}^{*}$, the  quasi-particle gap is opened instead by
the magnetization and Rashba SOI~\cite{PhysRevB.81.125318}, leading to
an  intrinsic AHE which  coexists with  topological superconductivity.
In this way, we find that  the intervening SHE at the quantum critical
point is correlated with the  change in the momentum-space topology of
the    BdG   quasiparticles    shown   in    the   lower    panel   of
Fig.~\ref{bdg_bands}.

As tuning $m_{z}$ through $m_{z}^{*}$ leads to a crossing of bands 
at four points in momentum space ($(k_{x},k_{y})=(\pm\pi,0)$ and 
$(0,\pm\pi)$), we can expect that the transition is first order in nature. As 
discussed by Volovik~\cite{volovik}, the topological nature of this 
Lifshitz transition is further revealed by computing the Chern invariant 
$\tilde{N}_{3}$ related to the  momentum space topology of the 
low-energy effective Dirac Hamiltonian, eq.(\ref{effcritHam}) (see supplementary 
information section C for details). 
We find that $\tilde{N}_{3}$ changes from $-\frac{1}{2}$ to 
$\frac{1}{2}$ as $m_{z}$ is tuned through $m_{z}^{*}$ from below, with 
the massless Dirac point at $m_{z}=m_{z}^{*}$ possessing a 
Chern invariant $N_{3} = \tilde{N}_{3}(m_{z}>m_{z}^{*}) - \tilde{N}_{3}(m_{z}<m_{z}^{*}) = 1$~\cite{volovik}. 
This corresponds to a singularity associated with a Dirac monopole 
in momentum space. Scattering an electron from just below one of the Dirac 
points to just above the other can be described by a
Berry phase-carrying instanton tunneling event that interpolates between the
ground state of the system and an excited state that lies vanishingly close.
The geometric phase 
associated with the massless Dirac electron observed at $m_{z}=m_{z}^{*}$ is $\pi$, while that 
with the instantons that tunnel between the ground state and the nearby 
excited state is $\pi/2$. As the Brillouin zone (i.e., the configuration space) connecting 
these two states is multiply connected, a destructive interference 
mechanism causes the tunnel splitting related to these instantons to vanish. 
This stabilizes the spin Hall ground state against the scattering of electrons between 
the Dirac cones. The 
spin Hall critical point is instead destabilised when the emergent time-reversal symmetry of the 
Dirac electrons is explicitly broken for $m_{z}\neq m_{z}^{*}$; this spoils the interference 
mechanism acting on the instantons, and generates a gap in the spectrum through a first 
order transition.
The role played by topological excitations makes clear that this phase transition falls 
outside the Ginzburg-Landau-Wilson (GLW)
paradigm. A field-theoretic description of this transition, the renormalisation 
group (RG) scaling relations and the $T=0$ RG phase diagram are presented 
in the supplementary materials section D.

\begin{figure*}[!ht]
\begin{center}
\epsfig{file=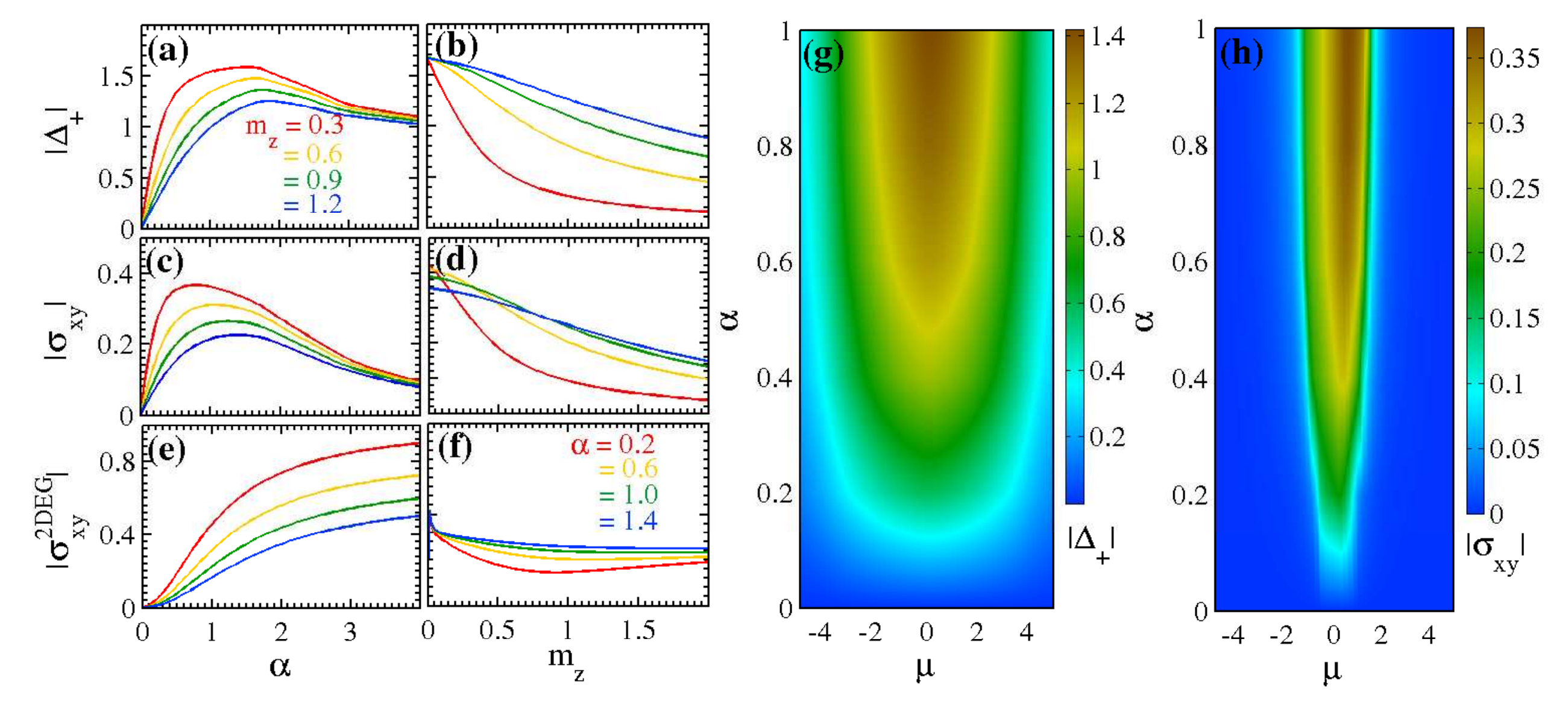, width=170mm}\vspace{-2em}
\caption{(Color  online) {\bf Dependence of Pairing amplitude and Hall conductances on other parameters.} 
  Figures  (a)-(f) shows  the variation  of the
  pairing   amplitude  $|\Delta_+|$   and   AHC  $|\sigma_{xy}|$   and
  $|\sigma_{xy}|^{2DEG}$ (in  units of $e^2/(2h)$)  with magnetization
  $m_z$ and Rashba  SOI $\alpha$ with $\mu=0$, $U=2.0$  and $t=1$. The
  non-monotonic feature  of $|\sigma_{xy}|$ (Fig.   (c)) distinguishes
  the novel  AHE from the  conventional AHE observed  in ferromagnetic
  Rashba  model without  superconductivity.  On  the other  hand, both
  $|\Delta_+|$ and AHC $|\sigma_{xy}|$ decays monotonically due to the
  explicit dependence in $\Delta$ which reduces with increasing $m_z$
  in the self-consistent calculation.   Please note that, in Fig.(d),
  $|\sigma_{xy}|$ is  valid except at $m_z=0$  where the time-reversal
  symmetry is intact and we get an emergent SHE.  Figures (g) and (h)
  are  the  plots  of  the  pairing  amplitude  $|\Delta_+|$  and  AHC
  $|\sigma_{xy}|$ (in  units of $e^2/(2h)$)  in the $\mu-\alpha$-plane
  for constant $U=2.0$, $m_{z}=0.5$  and $t=1$. This shows that although
  superconductivity can  be in a large filling-range,  the AHE appears
  only within a narrow window of filling.}
\label{alpha_hz}\vspace{-2em}
\end{center}
\end{figure*}
While  the results  presented  above are  robust  for the  case of  a
proximity-effect induced singlet  superconducting pairing, we employ a
self-consistent BdG formalism  (see supplementary information section E) for the
case  when the  pairing originates  from an  intrinsic superconducting
instability of the 2D electronic system. In  Fig.~\ref{alpha_hz}(a), 
(c)  and  (e) the  variations  of $|\Delta_+|$, $|\sigma_{xy}|$ and 
$|\sigma_{xy}^{2DEG}|$ (anomalous Hall conductivity for a ferromagnetic 2DEG  
with Rashba SOI) with various $m_z$ and $\alpha$ are  shown. Interestingly, both  
$|\Delta_+|$ and $|\sigma_{xy}|$  reveal non-monotonic  behaviour  with 
respect  to increasing Rashba  SOI strength $\alpha$. This is due to the 
enhanced  rate of spin-precession due to an increased $\alpha$ acts
as  a  dephasing  mechanism  for superconductivity, thus reducing the 
superconducting pairing  amplitude $\Delta$.   Also,  this non-monotonic
behaviour   is    the   striking   difference   from    the   AHE   in
Ferromagnetic-2DEG with Rashba  SOI (shown in Fig.~\ref{alpha_hz}(e)).
A point-contact  Andreev-tunneling spectroscopy  can be used  to probe
this new  type of AHE  in Rashba-coupled superconductor. On  the other
hand, both $|\Delta_+|$ and  $|\sigma_{xy}|$ are seen to decrease with
increasing $m_z$ as in FIG.~\ref{alpha_hz}(b)  and (d). This is due to
the fact that both the quantities depend explicitly on the pairing gap
$\Delta$, and a self-consistent  treatment of $\Delta$ reveals that it
decreases with an increasing $m_z$ due to pair-breaking processes. The
anomalous  hall  conductivity  $|\sigma_{xy}^{2DEG}|$  decreases  very
slowly  with  $m_z$  (shown  in Fig.~\ref{alpha_hz}(f))  because with
the increase of the gap at Fermi level, the Berry curvature reduces at very
slow  rate.  The  variations of  the  pairing gap  $\Delta_+$ and  AHC
$\sigma_{xy}$  with   the  chemical  potential  $\mu$   are  shown  in
Fig.~\ref{alpha_hz}(g)-(h).  The  pairing exists over  a large filling
range but the AHC is peaked only near the avoided level-crossings.

The nature of the variation of $|\Delta_+|$ with respect to the Rashba
SOI  ($\alpha$)  can shed  light  on  the  non-monotonic behaviour  of
superconductivity        observed        at        LaAlO$_3$/SrTiO$_3$
interface~\cite{Richter2013, Caviglia2008}.   In this Oxide interface,
superconductivity ($T_c\simeq 200$ mK) is observed at very low-filling
and the phase diagram is traced by varying the gate-voltage ($V_{g}$)
which  controls  both the  electron-concentration  ($n_{2D}$) and  the
Rashba   spin-orbit   splitting   ($\Delta_{so}$).   With   increasing
$n_{2D}$, the  Curie temperature  ($T_c$) should reveal  a dome-shaped
superconducting  phase   in  the  $n_{2D}-T_c$-space.  However,  the
enhanced  $\Delta_{so}$ serves as  a pair-breaking  agent and suppresses 
the superconductivity. With  increasing $V_{g}$, both $n_{2D}$ and $\Delta_{so}$ 
increase and the resulting competition between these  two opposing  effects 
leads to a  non-monotonic behaviour. 

Scattering from disorder, i.e, impurities which can be scalar potentials 
or even magnetic in nature, is a prominent feature of quantum transport.
For instance, while a Bardeen-Cooper-Schrieffer (BCS) superconductor  is  robust  against weak  
non-magnetic disorder ~\cite{Anderson195926}, sufficiently strong 
disorder can drive the system to a non-superconducting
state~\cite{dubinature2007}. It  is, therefore,  interesting to  analyse the
nature  of   the  AHE  and   its  coexistence  with   the  topological
superconductivity  in   the  disordered  situation.   The  topological
superconductivity  is  Rashba  SOI-generated  and  non-local. Disorder 
can give rise to additional contributions to the AHC from side-jump and 
skew-scattering mechanisms~\cite{RevModPhys.82.1539}. However, it 
cannot degrade the contribution from the intrinsic Berry phase mechanism 
described above. This is because the Chern invariant $\tilde{N}_{3}$ is a
topological property of a two-dimensional electronic system with broken 
time-reversal symmetry which was shown to be robust against electronic 
correlations and disorder scattering~\cite{ishikawamatsuyama}. We choose, 
therefore, to focus instead on the effects of disorder on the competing 
orders and the associated phase transition discussed above. 

We introduce non-magnetic disorder as a local shift in the chemical potential 
$H_{non-mag}=\sum_{i\sigma}V_d c_{i\sigma}^{\dagger} c_{i\sigma}$, where
$V_d$  is the  random disorder  potential which  varies  within [$-W$,
$W$], $W$ being the maximum strength of the disorder. 
\begin{figure*}[!ht]
\begin{center}
\epsfig{file=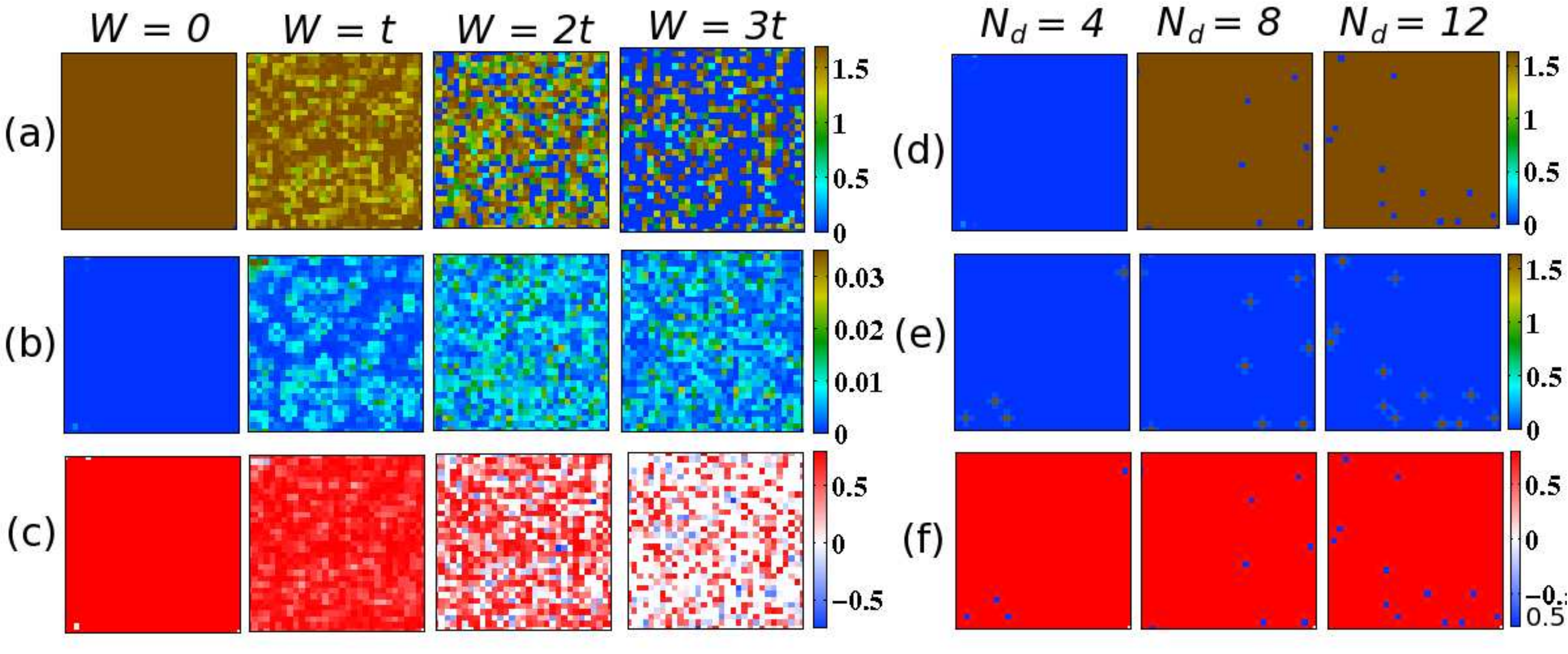, width=170mm}\vspace{0em}
\caption{(Color online) {\bf Real-space effects of disorder.} 
  The spatial  distribution of the local pairing
  amplitude $|\Delta(r_i)|$  (top row) and  magnetization (middle row)
  in  31$\times$31  square  lattice in the presence of disorder. The
  lowest row  shows the spatial  profile of the dominating  order; red
  and  blue   colours  represent  regions   of  superconductivity  and
  ferromagnetism. (a) - (c) The  columns  are  for  non-magnetic disorder
  strength $W  = 0$, $t$  , $2t$ and  $3t$ (from left to  right). 
  The parameters used  are $U=-1$,  $m_z =  0.5$ and
  $\alpha = 0.8$. (d) - (f) The columns are for different number of 
  magnetic impurities $N_d= 4$,  $8$ and $12$ (from left  to right). The
  parameters used are $U = -1$, $J_H = 1$ and $\alpha = 0.8$.}
\label{disorder_all}\vspace{0em}
\end{center}
\end{figure*} \vspace{0em}
FIG.~\ref{disorder_all} (a) - (c) represents  the  real-space profile of the pairing  gap
$\Delta(r_i)$   and the  magnetization   $m(r_i)$   calculated   via   a
self-consistent  BdG  formalism (see  supplementary
information section F).  In the strongly disordered system, these two competing
orders       stay        apart       in       spatially       excluded
regions~\cite{mohanta_phase_segregation}.   It  is, therefore,  inferred
that the AHE observed in ferromagnetic Rashba model resides in the 
ferromagnetic regions  while the novel AHE exists  in the  superconducting 
islands.  In  this way,  disorder enables  the mesoscale phase 
coexistence~\cite{dagotto} of the two types of AHE. Thus, the 
sharp phase boundary in the phase diagram 
Fig.~\ref{ahe_phase}(b) should be replaced with a phase separation 
region, with the transition 
belonging to the class of problems involving quantum percolation.
The presence of massless Dirac electrons carrying $\pi$ Berry phase 
was observed recently by Fernandes and Schmalian to be crucial for 
the appearance of complex critical exponents in the percolation 
transition in the 2D disordered Josephson junction 
array~\cite{fernandesschmalian};  clearly, such a transition falls 
outside the GLW paradigm. The same can be expected for 
the transition in the present problem with disorder. Further, when the 
Rashba  SOI and Zeeman field is large enough to  convert the ordinary 
superconductivity into  a topological one,  the  resulting topological 
superconductivity  will  be increasingly vulnerable  to  non-magnetic 
impurity  scattering due  to  explicitly broken time-reversal 
symmetry~\cite{PhysRevB.83.184520}.
%
Magnetic  impurities, on  the  other hand,  have  detrimental effects  on
$s$-wave  superconductors.   At   the  mean-field  level,  the  magnetic
impurities can be described by the Hamiltonian
$\hat{H}_{mag}=-J_{H}/2\sum_{j,\sigma,\sigma^{\prime}}       \boldsymbol{\sigma}_{\sigma
  \sigma^{\prime}}^{z}
c_{j\sigma}^{\dagger}c_{j\sigma^{\prime}}$,
where $j$ runs  over $N_d$ number of impurity  sites, randomly located
in  the  two-dimensional space  (see  supplementary information section G).   As
shown in FIG.~\ref{disorder_all} (d) - (f), superconductivity is totally destroyed at the
impurity  sites   and  we  again   have  the  spatial   separation  of
superconductivity and  ferromagnetism. As in the  case of non-magnetic
disorder, here also the two types of AHE coexist: the conventional AHE
resides at the impurity sites and the novel AHE at the superconducting
regions.

To summarize, we have proposed a novel AHE of BdG quasiparticles which
arises  from an  interplay of  singlet superconducting  pairing  and a
finite perpendicular  magnetization in presence  of Rashba SOI.   At a
critical  magnetization,  an emergent  spin  Hall  phase  is found  to
coincide    with   a   transition    from   normal    to   topological
superconductivity. The quantum phase transition is shown to be driven 
by instanton excitations that change the topological nature of the ground 
state.  At  yet  larger  magnetization,  the  topological
superconductivity  coexists   with  the   AHE.   For  the   case  when
superconductivity is suppressed completely by magnetization, we expect
a first order phase transition into an emergent intrinsic AHE phase of
the  ferromagnetic  2DEG  Rashba  model.   We  observed  non-monotonic
behaviour of  the AHC with respect  to the Rashba SOI  strength due to
the  fact that  at  very  large SOI,  the  enhanced spin-precession  is
pair-breaking for  superconductivity.  In this  connection, we explain
the   non-monotonic   feature   of   superconductivity   observed   at
LaAlO$_3$/SrTiO$_3$  interface   when  gate-voltage  is   tuned.   The
non-monotonic dependence of the AHC on Rashba SOI in the presence of a
superconducting order  parameter can  be easily  distinguished from
that obtained in its  absence.  The coexistence of superconductivity
and the  AHE is also  expected to lead  to a subtle  interplay between
Andreev   reflection   and  Hall   conductivity   of  edge-state   BdG
quasiparticles in  several of these phases. Additionally, we have
studied the  effects of disorder  in this model-system and  found that
both magnetic and  non-magnetic disorder result in a  coexistence of
the  novel AHE with  the conventional  AHE observed  in ferromagnetic
Rashba  model. The  scenario presented  here can likely also be observed in thin
film superconductors with broken time-reversal or 
centro-inversion symmetries. 

\section*{acknowledgement}
N.M.   acknowledges   MHRD,  India  for   support.   S.L.   gratefully
acknowledges support from the DST, Govt.  of India through a Ramanujan
fellowship. S.B.   and S.L.  thank CTS,  IIT-Kharagpur for hospitality
while a  part of the work  was conducted. S.B. and  S.L. thank Anirban
Mukherjee  for  several discussions and invaluable help  with  visualizing  
the Fermi  surface topology diagrams.

\section*{Author contributions}
All authors contributed equally to the development of the model, its analysis 
as well as the writing of the main text and supplementary materials.

\section*{Additional information}
The authors declare no competing financial interests. Supplementary information accompanies this paper. 
Correspondence and requests for materials should be addressed to S.L. and A. T.


\clearpage

\pagebreak
\widetext
\twocolumngrid
\begin{center}
\textbf{\large Supplementary Informations}
\end{center}
\setcounter{equation}{0}
\setcounter{figure}{0}
\setcounter{table}{0}
\setcounter{page}{1}
\makeatletter
\renewcommand{\theequation}{S\arabic{equation}}
\renewcommand{\thefigure}{S\arabic{figure}}
\renewcommand{\bibnumfmt}[1]{[S#1]}
\renewcommand{\citenumfont}[1]{S#1}

\subsection{Pairing symmetry in Rashba-split bands}
The  energy bands  (consider Hamiltonian  (1), in  main  text, without
superconductivity)  created by  the Rashba  SOI and  magnetization are
given  by  $\epsilon_{\pm}({\mathbf{k}})=\epsilon_k  \pm  \xi$,  where
$\xi=(\alpha^2|\mathbf{g_k}|^2+m_z^2)^{1/2}$   and  the  corresponding
eigenstates   $[c_{k,+},c_{k,-}]$  are   obtained  by   the  following
transformation
\begin{equation}
\begin{split}
\begin{pmatrix} \begin{array}{c} c_{k\uparrow} \\ c_{k\downarrow} \end{array} \end{pmatrix} &= \frac{1}{\sqrt{2}}\begin{pmatrix} \begin{array}{cc} a_1 & b_1e^{i\phi}\\a_2e^{-i\phi} & b_2 \end{array} \end{pmatrix} \begin{pmatrix} \begin{array}{c} c_{k,+} \\ c_{k,-} \end{array} \end{pmatrix}
\end{split}
\end{equation}
where $\phi=\tan^{-1}(\sin k_x /\sin k_y)$,
$a_1 a_2 = b_1 b_2 = -\alpha |\mathbf{g_k}|/(2\xi)$ and $a_1 b_2 - b_1 a_2 = m_z/\xi$.
When  written  in the  quasi-particle basis  $[c_{k,+},c_{k,-}]$,
Hamiltonian (1), in main text, reduces to 
\begin{equation}
\begin{split}
{\cal H}&=\sum_k [\epsilon_+({\mathbf{k}})c_{k,+}^{\dagger}c_{k,+}+\epsilon_-({\mathbf{k}})c_{k,-}^{\dagger}c_{k,-}\\
&+\Delta_{\pm} c_{k,\pm}^{\dagger}c_{-k,\pm}^{\dagger}+\Delta_s c_{k,+}^{\dagger}c_{-k,-}^{\dagger} + h.c.]
\end{split}
\label{Heff1}
\end{equation}
where $\Delta_{\pm}  = (-\alpha  |\Delta|/(2\xi))(\sin k_y \pm  i \sin
k_x)$  and  $\Delta_s  =  m_z  |\Delta|/\xi$  are,  respectively,  the
intra-band    and   inter-band    pairing    amplitudes.    Evidently,
$\Delta_{\pm}$ has chiral $p$-wave pairing symmetry whereas $\Delta_s$
is of $s$-wave symmetry.

\subsection{Calculation of Berry curvature of Bogoliubov-de Gennes bands}
Any 2x2 traceless Hamiltonian can be written as
\begin{equation}
H=\begin{pmatrix}f_3&f_1-if_2\\f_1+if_2&-f_3\end{pmatrix}~=~\sum_{i=1,2,3} f_i\sigma_i~
\label{fform},
\end{equation}
where the $\sigma_{i}$ are the three Pauli matrices. Now, the relation for the Berry curvature $\Omega$ can be written in the following form
\begin{equation}
\Omega =-\Im\frac{\langle+|\nabla_k H|-\rangle\times\langle-|\nabla_k H|+\rangle}{4E^2}~,
\end{equation}
where $E$ is the modulus of the eigenvalue of the Hamiltonian and $|\pm\rangle$ are the normalized eigenstates. In our case,
$E=\sqrt{f_1^2+f_2^2+f_3^2}~,~\nabla_k H=\sigma_i\nabla_kf_i$ and 
\begin{equation}
|\pm\rangle=\frac{1}{\sqrt{2E(E-f_3)}}\begin{pmatrix}f_1-if_2\\\pm E-f_3\end{pmatrix}~.
\end{equation}
As, $\nabla_kf_i$ is a number, we have to calculate the following matrix elements
\begin{eqnarray}
&&\langle-|\sigma_1|+\rangle=\frac{-f_1f_3+iEf_2}{E(E^2-f_3^2)}~,~\langle-|\sigma_2|+\rangle=\frac{-f_2f_3-iEf_1}{E(E^2-f_3^2)}~,\nonumber\\
&&\langle-|\sigma_2|+\rangle=\frac{\sqrt{E^2-f_3^2}}{E}~.
\end{eqnarray}
Now, the formula for Berry curvature $\Omega$ will be
\begin{equation}
\Omega =-\Im\frac{(\langle+|\sigma_i|-\rangle\nabla_kf_i)\times(\langle-|\sigma_i|+\rangle\nabla_kf_i)}{4E^2}~.
\end{equation}
First consider the term, $\nabla_kf_1\times\nabla_kf_2$ with a coefficient $c_3$
\begin{eqnarray}
c_3&=&\Im(\langle+|\sigma_1|-\rangle \langle-|\sigma_2|+\rangle-\langle+|\sigma_2|-\rangle\langle-|\sigma_1|+\rangle)\nonumber\\
&=&\frac{2Ef_3(f_1^2+f_2^2)}{E^2(E^2-f_3^2)}=\frac{2f_3}{E}~.
\end{eqnarray}
Now, using the rotational symmetry of the $\sigma$ matrices we get
\begin{eqnarray}
&&\Omega~=\nonumber\\ 
&& -\frac{f_3(\nabla_kf_1\times\nabla_kf_2)+f_2(\nabla_kf_3\times\nabla_kf_1)+f_1(\nabla_kf_2\times\nabla_kf_3)}{2E^3}~..
\end{eqnarray}

Now, the chiral bands in the proximity of the Fermi level are the $+$ bands, whose $2\times 2$ Hamiltonian is given by 
\begin{equation}
H_1=\begin{pmatrix}\epsilon_+&\Delta_+\\\Delta_+^*&-\epsilon_+\end{pmatrix}~~\text{for basis, }(c_{k+}^\dagger, c_{-k+})~,
\label{topSCham1}
\end{equation}
where $\epsilon_+=-2t(\cos k_x+\cos k_y)-\mu+\xi~$,~$\xi=\sqrt{\alpha^2(\sin^2k_x+\sin^2k_y)+m_z^2}~$,
~$\Delta_{+}=-\frac{\alpha|\Delta|}{2\xi}(\sin k_y+ i\sin k_x)$, with $\alpha$ being the Rashba SOI strength,$m_z$ is the 
magnetisation, $\Delta$ is the superconducting pairing gap, $\mu$ is the chemical potential and $t$ the hopping parameter. 
Mapping onto the general form of the $2\times 2$ matrix form given above in equation (\ref{fform}), the Berry curvature about 
the z-axis in spin space can be calculated from the relation
\begin{equation}
\Omega=\frac{-f_{3}(\partial_{k_y}f_1\partial_{k_x}f_2-\partial_{k_y}f_2\partial_{k_x}f_1)}{2E^3}
\end{equation}
where $f_{1}=\mathrm{Re}(\Delta_{+})=-\frac{\alpha|\Delta|\sin k_y}{2\sqrt{\alpha^2(\sin^2 k_x+\sin^2 k_y)+m_z^2}}$, 
\newline $f_{2}=\mathrm{Im}(\Delta_{+})\equiv\Delta_{-}=-\frac{\alpha|\Delta|\sin k_x}{2\sqrt{\alpha^2(\sin^2 k_x+\sin^2 k_y)+m_z^2}}$, \newline $f_{3}=\epsilon_{+}$ and $E=\sqrt{f_1^2+f_2^2+f_3^2}=\sqrt{\epsilon_+^2+\Delta_+\Delta_-}$. 
We can now compute the elements needed for the calculation of the Berry curvature about the z-axis 
\begin{eqnarray}\nonumber
\partial_{k_y}f_1&=&-\frac{\alpha|\Delta|\cos k_y}{2\sqrt{\alpha^2(\sin^2 k_x+\sin^2 k_y)+m_z^2}}\nonumber\\ 
&+&\frac{\alpha^3|\Delta|\sin^2 k_y\cos k_y} {2(\sqrt{\alpha^2(\sin^2 k_x+\sin^2 k_y)+m_z^2})^3}\\
\partial_{k_x}f_2&=&-\frac{\alpha|\Delta|\cos k_x}{2\sqrt{\alpha^2(\sin^2 k_x+\sin^2 k_y)+m_z^2}}\nonumber\\ 
&+&\frac{\alpha^3|\Delta|\sin^2 k_x\cos k_x} {2(\sqrt{\alpha^2(\sin^2 k_x+\sin^2 k_y)+m_z^2})^3}\\
\partial_{k_x}f_1&=&\frac{\alpha^3|\Delta|\sin k_y\sin k_x\cos k_x} {2(\sqrt{\alpha^2(\sin^2 k_x+\sin^2 k_y)+m_z^2})^3}\\
\partial_{k_y}f_2&=&\frac{\alpha^3|\Delta|\sin k_y\sin k_x\cos k_x} {2(\sqrt{\alpha^2(\sin^2 k_x+\sin^2 k_y)+m_z^2})^3}~.
\end{eqnarray}
From here, we compute
\begin{eqnarray}
&&\partial_{k_y}f_1\partial_{k_x}f_2=\\\nonumber
&&\frac{\alpha^6|\Delta|^2\sin^2k_x\sin^2k_y\cos k_x\cos k_y}{4(\alpha^2(\sin^2 k_x+\sin^2 k_y)+m_z^2)^{3}}\nonumber\\
&&+\frac{\alpha^2|\Delta|^2\cos k_x\cos k_y}{4(\alpha^2(\sin^2 k_x+\sin^2 k_y)+m_z^2)}\nonumber\\
&&\frac{\alpha^4|\Delta|^2(\sin^2k_x\cos k_x\cos k_y+sin^2k_y\cos k_x\cos k_y)}{4(\alpha^2(\sin^2 k_x+\sin^2 k_y)+m_z^2)^{2}}\\
&&\partial_{k_y}f_2\partial_{k_x}f_1=\frac{\alpha^6|\Delta|^2\sin^2k_x\sin^2k_y\cos k_x\cos k_y}{4(\alpha^2(\sin^2 k_x+\sin^2 k_y)+m_z^2)^3}\\
&&\partial_{k_y}f_1\partial_{k_x}f_2-\partial_{k_x}f_1\partial_{k_y}f_2\nonumber\\
&&=\frac{\alpha^2|\Delta|^2 m_z^2\cos k_x\cos k_y}{4(\alpha^2(\sin^2 k_x+\sin^2 k_y)+m_z^2)^2}~.
\end{eqnarray}
We can now finally compute the Berry curvature $\Omega$ as
\begin{eqnarray}
&&\Omega=(\partial_{k_{y}}f_{1}\partial_{k_{x}}f_{2}-\partial_{k_{y}}f_{2}\partial_{k_{x}}f_{1})\cdot f_{3}/(2E^{3})\nonumber\\
&&=\frac{\alpha^{2}|\Delta|^{2} m_{z}^{2}\cos k_{x}\cos k_{y}}{4(\alpha^{2}(\sin^{2} k_{x}+\sin^{2} k_{y})+m_{z}^{2})^{2}}\frac{\epsilon_{+}}{2(\epsilon_{+}^{2}+\Delta_{+}\Delta_{-})^{\frac{3}{2}}}\nonumber\\
&&=\frac{m_{z}^{2}\alpha^{2}|\Delta|^{2}\epsilon_{+}\cos k_{x} \cos k_{y}}{8\xi^{4}(\epsilon_{+}^{2}+\Delta_{+}\Delta_{-})^{\frac{3}{2}}}~.
\end{eqnarray}
This is the result presented in the main text.

\subsection{Derivation   of  effective   low-energy   Hamiltonian  and
  calculation  of  the  effective   velocity  of  the  emergent  Dirac
  quasi-particles}
A Rashba Hamiltonian in the spin ($\uparrow,\downarrow$) basis 
\begin{equation}
H_{R} = \begin{pmatrix}\epsilon_k-m_z&\alpha g_k\\\alpha g_k^*&\epsilon_k+m_z\end{pmatrix}
\end{equation}
can be diagonalized in the chiral basis ($+ \/ -$)
\begin{equation}
H_R=\begin{pmatrix}\epsilon_+&0\\0&\epsilon_-\end{pmatrix}~,
\end{equation}
where $\epsilon_\pm=\epsilon_k\pm\sqrt{\alpha^2|g_k|^2+m_z^2}~~~\&~~~g_k=\sin{k_y}+i\sin{k_x}$.
Upon including finite triplet pairing $\Delta_{\pm}$ and singlet pairing $\Delta_{s}$ order 
parameters, the Rashba Hamiltonian can be written in the Nambu basis, 
$(c_{k+},c_{k-},c^\dagger_{-k-},c^{\dagger}_{-k+})^T$, as 
\begin{equation}
H=\begin{pmatrix}\epsilon_+&0&\Delta_s&\Delta_+\\0&\epsilon_-&\Delta_-&\Delta_s\\\Delta_s&\Delta_-^*&-\epsilon_-&0\\\Delta_+^*&\Delta_s&0&-\epsilon_+\end{pmatrix}~,
\end{equation}
where the triplet pairing $\Delta_\pm=-\frac{\alpha|\Delta|}{2\sqrt{\alpha^2|g_k|^2+m_z^2}} (\sin k_{y} \pm i\sin k_{x})$ and the singlet pairing is $\Delta_s=\frac{m_z|\Delta|}{2\sqrt{\alpha^2|g_k|^2+m_z^2}}$.
This Hamiltonian can easily be projected to a 2x2 Hamiltonian in two cases: (i) triplet pairing is quite small in compared to the singlet pairing ($\Delta_{\pm}<<\Delta_{s}$) and (ii) the triplet pairing is much larger than the singlet pairing ($\Delta_{\pm}>>\Delta_{s}$). For the first case, the Rashba 
spin-orbit coupling $\alpha$ is very small and leads to small band mixing. This is not the case of interest in the present study, and henceforth, we will focus on the case (ii). Here, bands with similar chirality will undergo mixing. We will consider this case in four different situations, as discussed below.

\textbf{Case I: $m_{z}=0$}\\
Here, $\Delta_s$ goes to zero. The Hamiltonian then gets separated in two subspaces of different chirality
\begin{equation}
H_1=\begin{pmatrix}\epsilon_+&\Delta_+\\\Delta_+^*&-\epsilon_+\end{pmatrix}~~\text{for basis, }(c_{k+}^\dagger, c_{-k+})~,
\label{topSCham}
\end{equation}
and
\begin{equation}
H_2=\begin{pmatrix}\epsilon_-&\Delta_-\\\Delta_-^*&-\epsilon_-\end{pmatrix}~~\text{for basis, }(c_{k-}^\dagger, c_{-k-})~.
\end{equation}
Each of $H_1$ and $H_2$ will give rise to their own topological invariant, but have opposite chirality. Hence, the charge currents arising from the two chiralities will ameliorate one another. In fact, at the $(k_{x},k_{y}) = (\pm\pi,0)$ and $(0,\pm\pi)$ points, the charge current contributions from $H_1$ and $H_2$ are identical in magnitude but opposite in sign; they will thus cancel one another. Thus, instead of an anomalous charge Hall conductivity, we find a spin Hall conductivity at these points in momentum-space.

\textbf{Case II: $m_{z}<<\alpha,\Delta$}\\
A small magnetization $m_{z}$ favours one of the chiralities and diminish transport effects due to the other. Further, the s-wave SC order parameter $\Delta_{s}$ causes mixing between the $+$ and $-$ chiralities and leads to a suppression of the anomalous Hall conductivity (AHC) in this regime.

\textbf{Case III: $\alpha>>m_{z}>>\Delta$}\\
As $m_z$ crosses the strength of $\Delta$, the system enters into the topological superconductor phase and only one chirality remains 
with the Hamiltonian given by eq.(\ref{topSCham}). We can see from this Hamiltonian that the mass for the Dirac electrons is given by 
the magnetization $m_{z}$ and the Rashba SOI coupling $\alpha$ (but not the SC order parameter $\Delta$). In this case we have anomalous Hall coefficient for only the $+$ chiral band, and hence find a higher value of the AHC here as compared to case II.

\textbf{Case IV: $m_{z}\simeq m_{z}^{*}$ }\\
The effective low-energy $2\times 2$ Hamiltonian is derived from an expansion of the full $4\times 4$ Hamiltonian around $(k_{x},k_{y}) = (\pm\pi,0)$ and $(0,\pm\pi)$ and $m_{z}\simeq m_{z}^{*}$. We choose $k_{x}$ and $k_{y}$ as small deviations ($k$) from $0$ and $\pi$ for convenience in the 
calculation. Here the eigenvalues of the two bands closest to the chemical potential are $E_{1}$ and $E_{2}=-E_{1}$ given by
\begin{eqnarray}
&&E_{1}=\nonumber\\
&&\sqrt{m_z^2+\Delta^2+\mu^2+\alpha^2k^2-2\sqrt{(\Delta^2+\mu^2)m_z^2+\mu^{2}\alpha^2k^2)}}\nonumber\\
&&=\sqrt{m_z^2+m_z^{*2}+\alpha^2k^2-\sqrt{m_{z}^{2}(m_{z}^{*2}+2\mu^{2}\alpha^2k^2)}}~,
\end{eqnarray}
where $m^*_z=\sqrt{\Delta^2+\mu^2}$ and we have taken $\sin k \sim k$ to lowest order. Observe that for $m_z\rightarrow m_z^*, k\rightarrow 0$, $E_{1}=0=E_{2}$, i.e., the band gap vanishes. Thus, the ``mass" term of the effective $2\times 2$ Hamiltonian can be written as $(m_{z}-m_{z}^{*})\sigma_{z}$ (to lowest order in $m_{z}-m_{z}^{*}$). To obtain the other terms in the effective $2\times 2$ Hamiltonian, we write the energy $E_{1}$ as
\begin{eqnarray}
E_{1}&=&\sqrt{m_{z}^{2}+m_{z}^{*2}+\alpha^{2}k^{2}-2m_{z}m_{z}^{*}\sqrt{1+(\frac{\mu\alpha k}{m_{z}m_{z}^{*}})^{2}}}\nonumber\\
&\simeq& \sqrt{m_z^2+m_z^{*2}+\alpha^2k^2-2m_{z}m_{z}^{*}(1+\frac{\mu^{2}\alpha^{2}k^{2}}{2m_{z}^{2}m_{z}^{*2}})}\nonumber\\
&=& (m_z^2+m_z^{*2}-2m_{z}m_{z}^{*} +\alpha^{2}k^{2}(1-\frac{\mu^{2}}{m_{z}m_{z}^{*}}))^{1/2}\nonumber\\
&=& (1-\frac{\mu^{2}}{m_{z}^{*2}})^{1/2}\alpha k~~~~\mathrm{for}~~~m_{z}=m_{z}^{*}~. 
\end{eqnarray}
This gives the velocity of the Dirac electrons as \newline $v=\alpha(1-\frac{mu^{2}}{m_{z}^{*2}})^{1/2}$. In this way, 
we find the effective $2\times 2$ Hamiltonian can be written as shown in the main text
\begin{equation}
H_{+}(\mathbf{k})=(v\sin k_{y})\sigma_{x}+(v\sin k_{x})\sigma_{y}+(m_{z}^{*}-m_{z})\sigma_{z}~.
\end{equation}

\subsection{The topological nature of the ground states, the excitations and the quantum phase transition}
\begin{figure*}[!ht]
\begin{center}
\includegraphics[width=170mm]{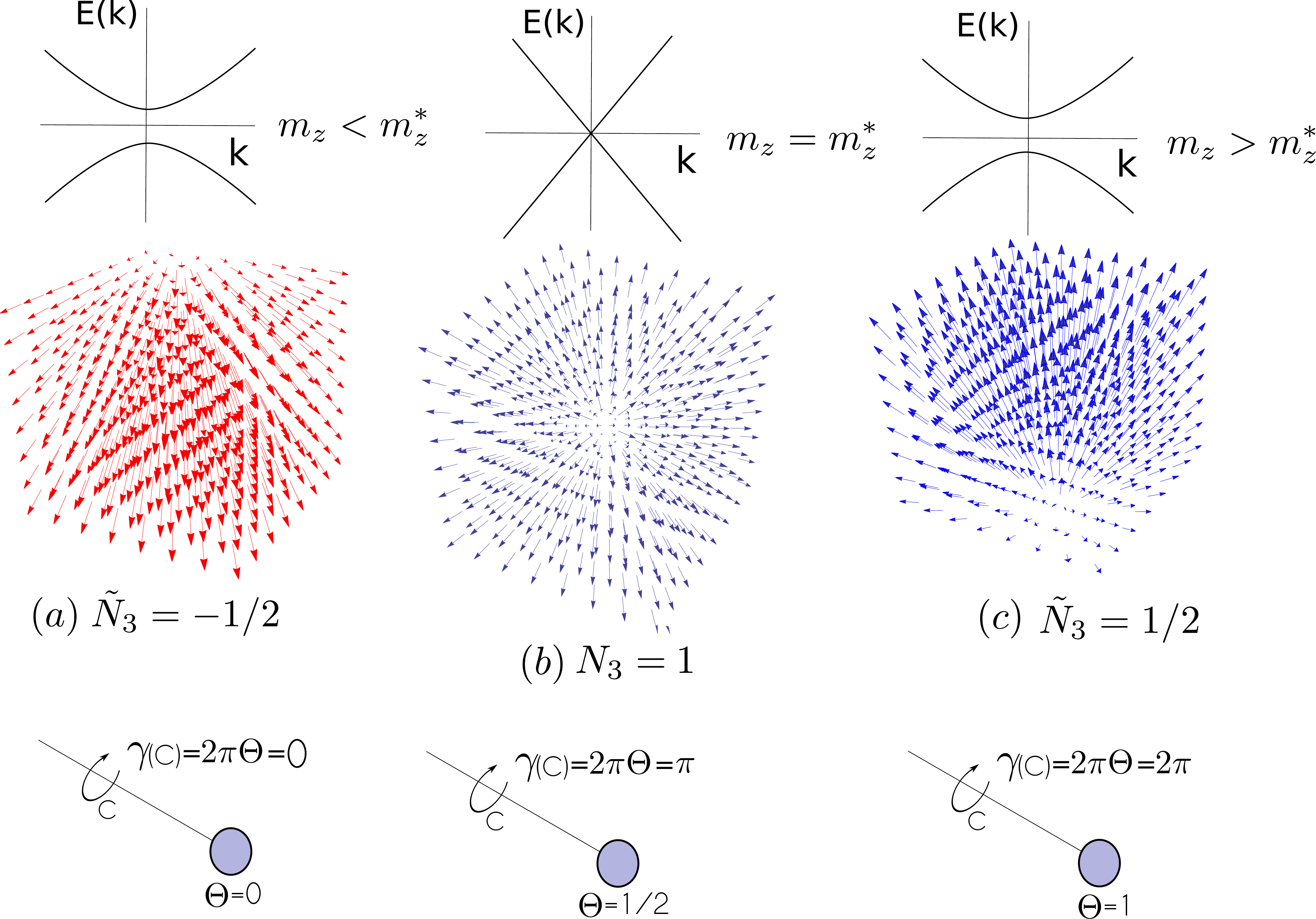}
\caption{(Color online) The topological Chern invariant associated with the momentum space structure 
of the massive Dirac equation as $m_{z}$ is tuned through $m_{z}^{*}$. (a) For $m_{z}<m_{z}^{*}$, the 
massive Dirac spectrum possesses a Chern invariant $\tilde{N}_{3}=-1/2$ corresponding to a 
half-hedgehog (or ``meron hedgehog") in momentum space. This topological defect is equivalent to a 
magnetic monopole of strength $\Theta=0$, with a Dirac string around which a Berry phase 
$\gamma = 2\pi\Theta = 0$ is accumulated in the circuit $C$. (b) For $m_{z}=m_{z}^{*}$, the 
massless Dirac spectrum has a Chern invariant 
$N_{3} = \tilde{N}_{3}(m_{z}>m_{z}^{*}) - \tilde{N}_{3}(m_{z}<m_{z}^{*})=1$ corresponding to a hedgehog 
in momentum space. This topological defect is equivalent to a magnetic monopole of strength 
$\Theta=1/2$, with a Dirac string around which a Berry phase $\gamma =\pi$ is accumulated in the 
circuit $C$. (c) For $m_{z}>m_{z}^{*}$, the massive Dirac spectrum possesses a Chern invariant 
$\tilde{N}_{3}=-1/2$ corresponding to an anti-half-hedgehog (or ``anti-meron hedgehog") in 
momentum space. This topological defect is equivalent to a magnetic monopole of strength 
$\Theta=1$, with a Dirac string around which a Berry phase $\gamma = 2\pi$ is accumulated in the 
circuit $C$.}
\label{hedgehogs}\vspace{-2em}
\end{center}
\end{figure*}

We begin this section by discussing briefly the topological nature of the Dirac 
Hamiltonian, as shown by Volovik~\cite{volovik}. 
We consider here a generic Hamiltonian for a system of 2+1D massive Dirac electrons
\begin{equation}
H= v(\sigma_{x}p_{x} + \sigma_{y}p_{y} + m\sigma_{z}/v)~,
\end{equation}
where $v$ is the velocity of the Dirac electrons and $m$ their mass. This eigenspectrum has two 
branches $E=\pm v\sqrt{p_{x}^{2}+p_{y}^{2}+m^{2}}$. The two branches are separated by the mass 
$m$ and touch each other at $m=0$. It is then possible to define a topological Chern invariant for 
this system~\cite{volovik}
\begin{equation}
\tilde{N}_{3}=\frac{1}{24\pi^{2}}\epsilon_{\mu\nu\lambda}\mathbf{Tr}\int d^{2}pdp_{0} G\partial_{p_{\mu}}G^{-1}~G\partial_{p_{\nu}}G^{-1}
~G\partial_{p_{\lambda}}G^{-1}~,
\end{equation}
where $p_{0}\equiv E$ (the energy), the Matsubara Greens function is given by
\begin{equation}
G = (ip_{0} - H(\vec{p},m))^{-1}~,
\end{equation}
and the trace $\mathbf{Tr}$ is taken over all eigenstates of the system. We can see immediately that 
the $G$ has a singularity at $(p_{0}=0,p_{x}=0,p_{y}=0,m=0)$: the Dirac point for the case of the 
massless Dirac spectrum. For the case of a finite, non-zero mass $m$, the system is gapped and the 
Greens function is regular everywhere in 3-momentum space $p_{\mu}=(p_{0},p_{x},p_{y})$. For the 
case of $m<0$ (which corresponds to $m_{z}<m_{z}^{*}$ in the original electronic problem), one finds 
$\tilde{N}_{3}=-1/2$ while for $m>0$ ((which corresponds to $m_{z}>m_{z}^{*}$), one finds 
$\tilde{N}_{3}=+1/2$. These two cases correspond to the existence of half-hedgehog topological 
defects in 3-momentum space: for $m<0$, this is the ``anti-meron-hedgehog" and for $m>0$, the 
``meron hedgehog" (see Fig.(\ref{hedgehogs}(a)) and Fig.(\ref{hedgehogs}(c)) respectively). These can 
equivalently be thought of as magnetic monopoles in the 3-momentum space with strength 
$\Theta=0,1$ respectively; the Berry phase accumulated by making a close circuit around the Dirac 
string attached to this magnetic monopole is given by $\gamma = 0,2\pi$ respectively.
 
For the case of the massless Dirac spectrum, i.e., $m=0$, the topological Chern invariant which 
describes the singular point at the origin of 3-momentum space corresponds to a winding number of 
the mapping of the spherical surface $S_{2}$ 
of $(\sigma_{x},\sigma_{y},\sigma_{z})$ around the singular point onto a 2-sphere of a unit vector 
$\hat{n}$~~\cite{volovik}
\begin{equation}
N_{3} = \frac{1}{8\pi}\epsilon_{ijk}\int_{S_{2}} \hat{n}\cdot(\frac{\partial\hat{n}}{\partial p_{i}}\times\frac{\partial\hat{n}}{\partial p_{j}})~.
\end{equation}   
For the case of massless 2+1D Dirac electrons, one finds $N_{3} =1$, corresponding to a hedgehog 
topological defect in 3-momentum space (see Fig.(\ref{hedgehogs}(b))). This can equivalently be 
thought of as a magnetic monopole in the 3-momentum space with strength $\Theta=1/2$; the Berry 
phase accumulated by making a close circuit around the Dirac string attached to this magnetic monopole 
is given by $\gamma = \pi$. Further, 
the two Chern invariants $N_{3}$ and $\tilde{N}_{3}$ are related to one another as~\cite{volovik}
\begin{equation}
N_{3} (m=0) = \tilde{N}_{3} (m>0) - \tilde{N}_{3} (m<0) = 1~.
\end{equation}

It is also important to note that for $m=0$, the massless 2+1D Dirac Hamiltonian 
$H=v(\sigma_{x}p_{x} + \sigma_{y}p_{y})$ is invariant under a time-reversal transformation 
$\sigma_{\mu}\rightarrow -\sigma_{\mu}$, $p_{\mu}\rightarrow -p_{\mu}$ $(\mu=x,y)$. 
This corresponds to an emergent time-reversal symmetry of the massless Dirac electrons 
observed at $(k_{x},k_{y}) = (\pm\pi,0)$ and $(0,\pm\pi)$ in our electronic problem, and should not be 
confused with the fact that the presence of a finite magnetisation $m_{z}$ means that the 
fundamental time-reversal symmetry of the system (related to the electron spin) is already 
broken. However, the emergent time-reversal symmetry (TRS) for the Dirac electron has an interesting consequence:
Kramers theorem~\cite{kramers} states that there must be a partner for such TRS electrons. 
In our problem, this corresponds to the emergence of pairs of gapless Dirac spectra, i.e., at 
$(k_{x},k_{y}) = (\pm\pi,0)$ and at $(k_{x},k_{y}) = (0,\pm\pi)$. Note that for a Hamiltonian $H$ with TRS invariance, there 
exists an antiunitary operator $T$ which commutes with $H$, $[H,T]=0$. Further, for 
an eigenstate $|\psi_{1}\rangle$ of $H$ whose energy is $E_{1}$, the commutator of $H$ 
and $T$ guarantees the existence of another eigenstate of $H$, $|\psi_{2}\rangle$, with 
the same energy: this results from $T|\psi_{1}\rangle = \pm|\psi_{2}\rangle$~\cite{kramers}. 
$|\psi_{1}\rangle$ and $|\psi_{2}\rangle$ are then referred to as a Kramers doublet. This 
has an important consequence for a spin Hall topological insulator state in the form of 
a pair counter-propagating helical edge states at every boundary of the system~\cite{sinova,kanemele}.
Further, the existence of two degenerate zero energy states (the 
Dirac points) in momentum space needs careful consideration, as a study of the 
electronic scattering processes between them is key towards understanding the phase 
transition as $m_{z}$ is tuned through $m_{z}^{*}$. This is carried out below.

From the above discussion, we can see that the topological feature of the ground 
state of the low-energy effective Dirac Hamiltonian is associated with topological 
textures in the spinorial structure of the momentum space of this problem. In order to 
understand the process by which this topological feature can change through a phase 
transition, as well as the excitations that drive the transition, we need to  examine the 
appearance of the pairs of gapless Dirac spectra at $(k_{x},k_{y})=(\pm\pi,0)$ and 
$(0,\pm\pi)$ in the bandstructure and the 
scattering processes that may connect them. Given that the momentum difference 
required for such scattering is large, we may expect that it can only be significant 
for the case of sufficiently strongly repulsive short ranged interactions between the 
massless Dirac electrons. For weakly repulsive interactions, we need to 
establish their relevance for the low-energy physics through renormalisation group (RG)
scaling transformations. This gives insight into whether the degeneracy 
of the two pairs of Dirac nodes can be lifted, and whether gaps can be opened in the  
Dirac spectra. The problem is further complicated by the fact that these Dirac 
quasiparticles carry Berry phases, as also seen from the discussion of the topological 
invariants above.

\begin{figure*}[!ht]
\begin{center}
\includegraphics[width=170mm]{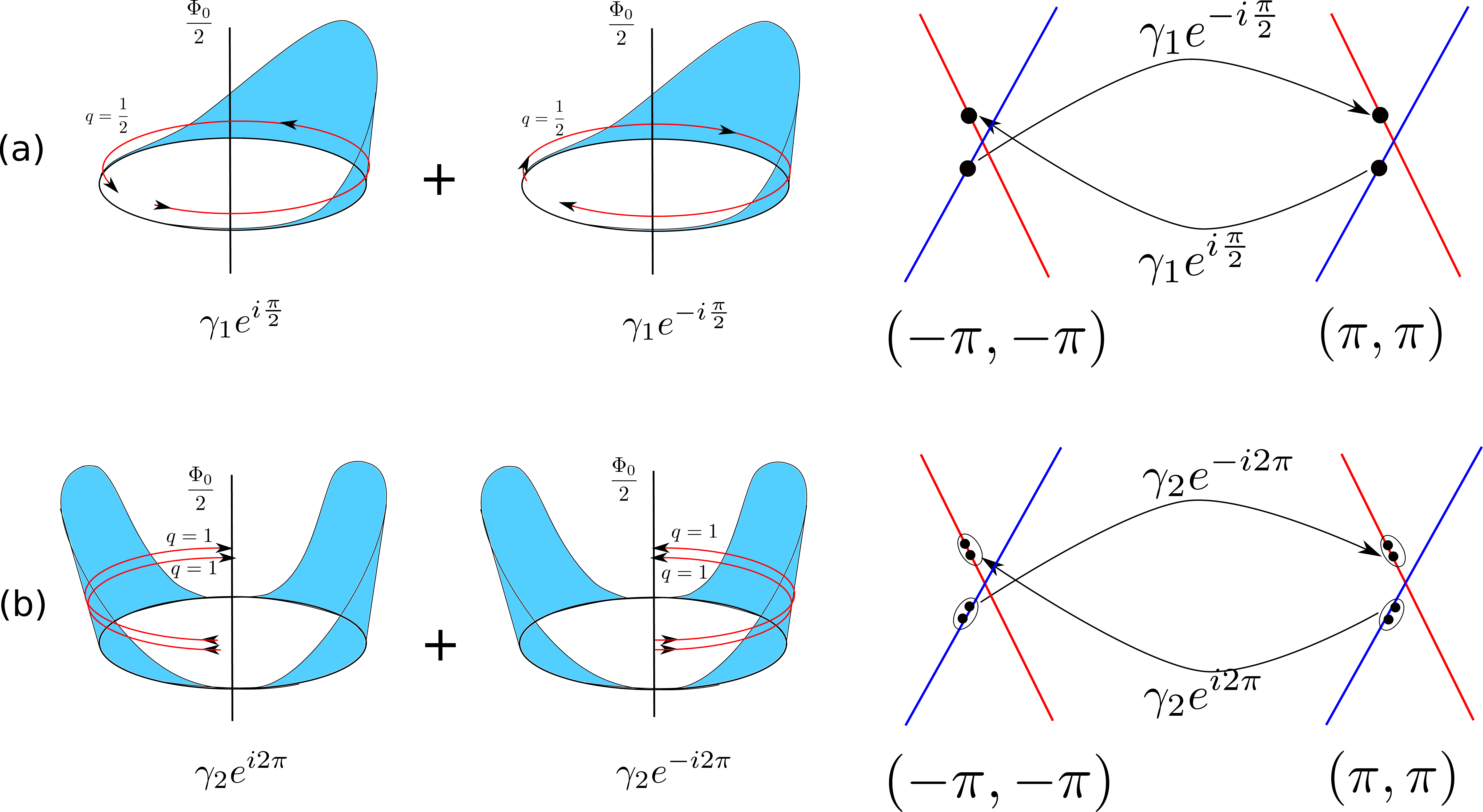}
\caption{
(Color  online) Single-particle backscattering and 2-particle Umklapp scattering between the 
two Dirac cones and their corresponding instanton tunneling processes. In (a), we show that 
the single particle backscattering is equivalent to an instanton event in which a particle with 
(topological) charge $q=1/2$, confined on a circle 
(characterised by an angle $\phi$) with a $\cos\phi$ potential, tunnels from $\phi=0$ to 
$\phi=2\pi$ (and vice versa) with tunnel amplitude $\gamma_{1}$. The circle is threaded by an 
Aharanov-Bohm (AB) flux $\Phi_{0}/2$, such that the instanton and anti-instanton pick up AB 
phases of $e^{i\pi/2}$ and $e^{-i\pi/2}$ respectively during the tunneling event. In (b), 
we show that the Umklapp scattering process is equivalent to an instanton event in which 2 particles, 
each with (topological) charge $q=1$, confined on a circle (characterised by an angle $\phi$) with 
a $\cos2\phi$ potential tunnels from $\phi=0$ to $\phi=\pi$ (and vice versa) with tunnel 
amplitude $\gamma_{2}$. The circle is threaded by an Aharanov-Bohm (AB) flux $\Phi_{0}/2$, 
such that the instanton and anti-instanton pick up AB phases of $e^{i2\pi}$ and $e^{-i2\pi}$ 
respectively during the tunneling event.
}
\label{diracscatt}\vspace{-2em}
\end{center}
\end{figure*}

The answers to these questions is achieved by constructing a non-linear sigma model 
(NLSM) theory that describes the long-wavelength, low-energy dynamics of the order 
parameter field describing the scattering of quasiparticles within each of the two pairs of 
Kramers doublet Dirac cones at $(k_{x},k_{y})=(\pm\pi,0)$ and $(0,\pm\pi)$. In this, we 
are guided by Senthil and Fisher~\cite{senthilfisher}, who following the classic work of 
Abanov and Wiegmann~\cite{abanovwiegmann}, showed that the $N=2~\mathrm{QED}_{3}$ 
theory in 2+1D is equivalent to the $\mathrm{O}(4)$ NLSM 
in 2+1D with a topological $\Theta$-term and a value of the topological angle 
$2\pi\Theta=\pi$. For the sake of completeness, we sketch briefly their approach and 
results here, before proceeding to study the phase transition through an alternate 
approach involving the non-trivial edge states of the system. We start by identifying 
the low-energy theory of a given Kramers doublet pair of Bogoliubov-de Gennes (BdG) 
quasiparticles with Dirac dispersion as a $N=2~\mathrm{QED}_{3}$ 
theory in Euclidean 3D, consisting of 2 ``flavours" of two-component Dirac fermions 
$\psi$ coupled to a non-compact $U(1)$ gauge field $A$:
\begin{equation}
S=\int d^{3}x \bar{\psi}(\tau_{i}(-i\partial_{i} - A_{i}))\psi + \frac{1}{2e^{2}}(\epsilon_{ijk}\partial_{j}A_{k})^{2}~,
\end{equation} 
where the $\tau_{i}$ are the Pauli matrices. A similar theory can be constructed for the 
other Kramers doublet pair as well. This action has a global $SU(2)\times SU(2)$ symmetry, 
where each of the two $SU(2)$ reflects on an invariance associated with unitary rotations that 
transform one of the two Dirac fermion flavours of a given Kramers doublet into the other. 
We then focus on scattering processes involving both flavours of Dirac electrons within each 
Kramers doublet. Such processes describe the separate ordering tendencies of the two $SU(2)$ 
vectors, each of which can be written as $\bar{\psi}\sigma\psi$ (where 
$\sigma$ characterises the 2-component ``flavour" space). For a given Kramers doublet, this 
can be introduced via a short-ranged repulsive inter-flavour interaction term
\begin{equation}
S_{\mathrm{int}} = \int d^{3}x \Lambda (\bar{\psi}\sigma\psi)^{2}~~,~~\Lambda>0~.
\end{equation}   
Now, employing a Hubbard-Stratonovich transformation, we can decouple this four-fermion 
interaction in terms of a $O(3)$-symmetric order parameter vector field $\vec{N}$ whose 
magnitude is the mass ($m_{e}>0$) that can be generated for the Dirac quasiparticles by the 
2-particle scattering processes (shown in Fig.(\ref{diracscatt}(a))),
$\vec{N}=m_{e}~\vec{n}=m_{e}~(\phi_{1},\phi_{2},\phi_{3})$, $\vec{n}^{2}=1$:
\begin{equation}
\frac{m_{e}^{2}\vec{n}^{2}}{2\Lambda} + i~m_{e}\vec{n}\cdot\bar{\psi}\vec{\sigma}\psi~.
\end{equation}
Note that the first term is a constant with no dynamics as $\vec{n}^{2}=1$. Thus, 
we write the total action as:
\begin{equation}
S=\int d^{3}x \bar{\psi}(\tau_{i}(-i\partial_{i} - A_{i}) + i~m_{e}\vec{n}\cdot\vec{\sigma})\psi + \frac{1}{2e^{2}}(\epsilon_{ijk}\partial_{j}A_{k})^{2}~.
\end{equation} 

By following the technique developed by Abanov and Wiegmann~\cite{abanovwiegmann}, one 
can integrate out the fermion fields by computing the fermionic determinant and carrying out 
an expansion in $1/m_{e}$, as well as integrate out the gauge field $A$. Senthil and 
Fisher~\cite{senthilfisher} show that the resulting theory 
is the $\mathrm{O}(4)$ NLSM in 2+1D, but with a $\Theta$-term which describes 
the topological defects of this theory. This NLSM is written in terms of a field 
$U$ which depends on a 4-component unit vector (all of whose elements are real) 
$\vec{\phi} = (\phi_{0},\phi_{1},\phi_{2},\phi_{3})$ on the three dimensional 
spherical surface ($S^{3}$). $U$ is an element of the compact $\mathrm{SU}(2)$ 
Lie group. The fact that this Lie group has a nontrivial homotopy group $\pi_{3}[\mathrm{SU}(2)] = Z$ 
(corresponding to the integer number of coverings of space-time coordinate space by the $\vec{\phi}$ 
field configuration) signals the presence of topological defect excitations of this NLSM theory.
The charge of these topological defects reflects on the Chern invariant/Berry phase $\tilde{N}_{3}$ 
of the 2+1D Dirac electrons discussed earlier. The component $\phi_{0}$ characterises valence 
bond (VBS/dimer) order in the system and relates to the Umklapp backscattering process of a 
pair of electrons between the two Dirac cones. This is shown in Fig.(\ref{diracscatt}(b)). Importantly, 
$\phi_{0}$ also characterises the 
half-hedgehog (or ``meron-hedgehog"~\cite{senthilfisher}) topological defects: the sign of 
$\phi_{0}$ relates to the two values of $\tilde{N}_{3}=\pm 1/2$. Further, the dynamics of 
$\phi_{0}$ opens the door towards changes in this topologically invariant quantity through the 
generation of hedgehog instanton excitations (with topological charge $N_{3}=1$).
\begin{figure}[!ht]
\vspace{-0em}
\begin{center}
\epsfig{file=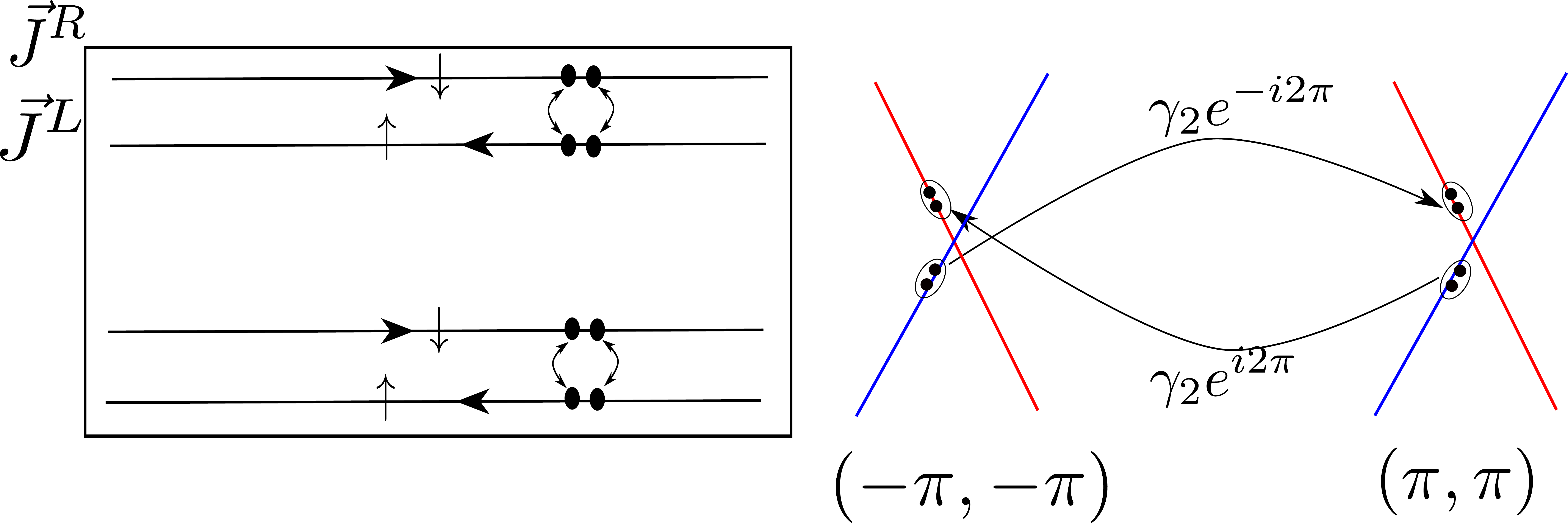, width=80mm}\vspace{-1em}
\caption{(Color online) Umklapp scattering of 2 electrons between the 2 oppositely directed helical edges of the 
spin Hall phase in our system.}
\label{edgescatt}\vspace{-1.5em}
\end{center}
\end{figure}\vspace{-0em}

The $\mathrm{O}(4)$ NLSM theory in 2+1D~\cite{senthilfisher,xuludwig} written in terms 
of the 4-component field $\vec{\phi}$ can be written as $S=S_{0} + i2\pi\Theta Q$~, where
\begin{equation}
 S_{0}=\int d^{2}x d\tau \frac{1}{g}(\partial_{\mu}\vec{\phi})^{2}~,
\end{equation}
and the topological charge $Q$ of the meron-hedgehogs is given by 
\begin{equation}
Q = \frac{1}{12\pi^{2}}\int d^{2}x d\tau \epsilon_{\mu\nu\rho}
 \epsilon_{abcd}\phi^{a}\partial_{\mu}\phi^{b}\partial_{\nu}\phi^{c}\partial_{\rho}\phi^{d}~.
\end{equation}  
The topological Chern invariant $\tilde{N}_{3}$ obtained earlier from the low-energy 
effective Dirac Hamiltonian is, in fact, equivalent to the topological charge $Q$ of the
meron-hedgehog defect configurations in the NLSM theory. The topological angle $2\pi\Theta = \pi$ 
for the case of massless Dirac electrons in the original electronic problem 
(i.e., $m_{z}=m_{z}^{*}$), while a finite mass for the Dirac electrons achieved for $m_{z}<m_{z}^{*}$ 
and $m_{z}>m_{z}^{*}$ corresponds to $2\pi\Theta<\pi$ and $2\pi\Theta>\pi$ respectively. 
In this way, the system undergoes a Lifshitz transition in which the topology of the electronic 
momentum space is changed as $m_{z}$ is tuned through $m_{z}^{*}$~\cite{lifshitz,volovik}.
At $2\pi\Theta=\pi$, the $T=0$ path integral of the system enjoys a discrete symmetry under 
a parity transformation of the four-vector $\vec{\phi}\rightarrow -\vec{\phi}$; this corresponds to 
an invariance under (Euclidean) time reversal transformation $i\tau \rightarrow -i\tau$ together 
with a parity transformation $(x,y,z)\rightarrow (-x,-y,-z)$. At this value of $\Theta$, 
one can equally as well see this as an invariance of the system under the parity transformation 
$\Theta\rightarrow -\Theta$; this is also true of the trivial values of $2\pi\Theta=0, 2\pi$. This 
discrete symmetry protects the topological angle from changing under RG transformation at these 
three values of $2\pi\Theta=0,\pi,2\pi$, while its flow under RG for all other values is not forbidden.   
At this point, we note that the appearance of the topological $\Theta$-term in the action give rise 
to the possibility of quantum interference in the free-energy landscape from which the quantum 
partition function is constructed. This arises from the complex-valued phase factors associated 
with different configurations of the system. It also shows that the true quantum many-body ground state 
is achieved via non-perturbative instanton tunneling events between various perturbative ground states 
which possess fixed values of the topological charge $Q$ and are weighted by $\Theta$-dependent 
phase factors. Such linear-superposition quantum ground states are called $\Theta$-vacua, 
and give rise to an oscillatory dependence of the energy separation between the ground state and the 
lowest lying excited state on the topological angle $\Theta$~\cite{rajaraman}.

\begin{figure*}[!ht]
\begin{center}
\includegraphics[width=150mm]{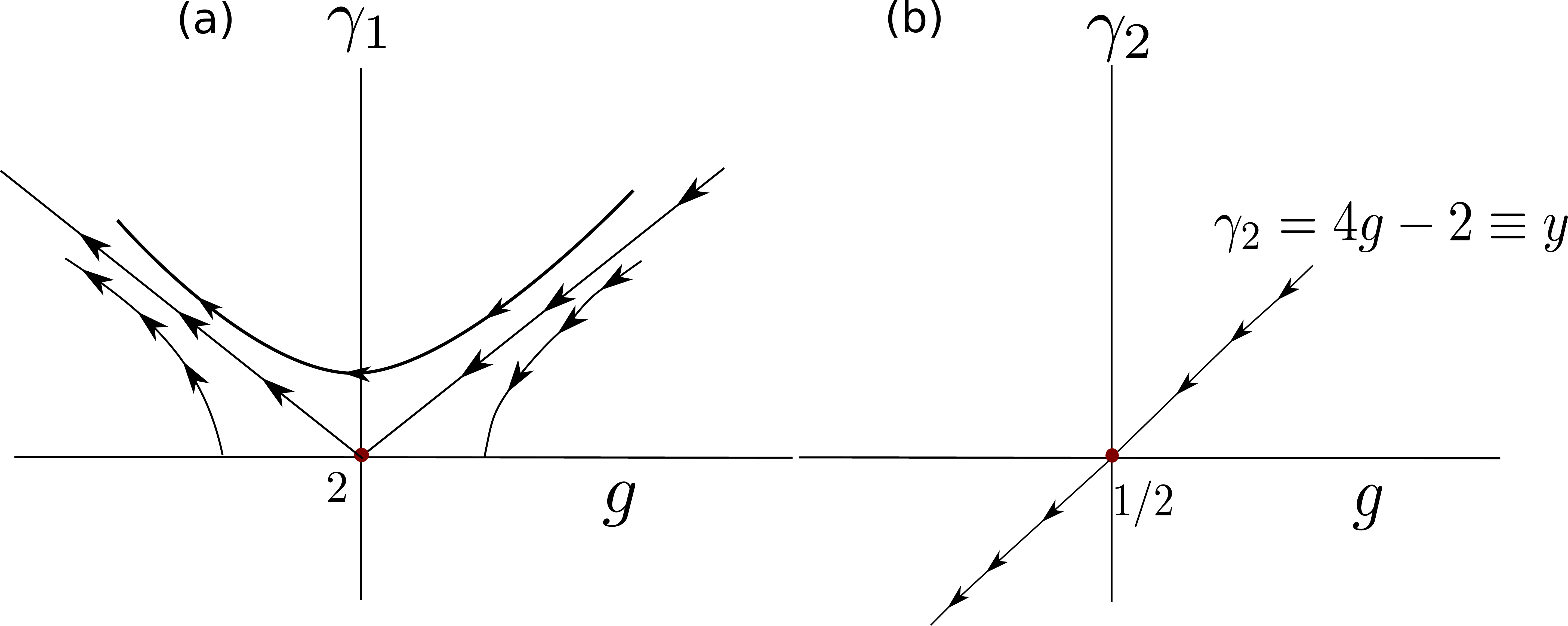}
\caption{(Color  online) The RG phase diagrams for the Berezinskii-Kosterlitz-Thouless (BKT) and 
Wess-Zumino-Novikov-Witten (WZNW) phase transitions. The diagram (a) is the BKT RG phase diagram 
displayed in terms of the NLSM coupling $g$ versus the single instanton tunnel amplitude $\gamma_{1}$. 
The separatrices $|g|=\gamma_{1}$ separate the RG trajectories leading towards the line of weak coupling 
fixed points $\gamma_{1}=0$, $g\geq 2$ from those leading towards the strong coupling fixed point in 
$\gamma_{1}$. The diagram (b) is the WZNW RG phase diagram displayed in terms of the NLSM coupling 
$g$ versus the doubled instanton tunnel amplitude $\gamma_{2}$. The RG flow on the $SU(2)$ symmetric 
line $\gamma_{2}=4g-2$ indicates the WZNW phase transition, with $\gamma_{2}$ being a dangerously 
irrelevant coupling under the RG transformations: for $\gamma_{2}>0$, the RG flow leads towards the 
$SU(2)$ symmetric weak-coupling fixed point theory at $g=1/2$, $\gamma_{2}=0$ (with a unique ground 
state) while for $\gamma_{2}<0$, the RG flow leads towards the strong coupling fixed point theory (with 
a doubly-degenerate ground state).}
\label{bktwznwrg}\vspace{-2em}
\end{center}
\end{figure*} 
The production and proliferation of such instantons, therefore, needs careful consideration. 
For instance, the topological defect instanton excitations that can change the value of a given $Q$ 
are expected to be suppressed by a large mass for the Dirac electrons. Should this be the case, 
we may expect to 
reach conclusions from studying only the kinetic part of the NLSM action. This is achieved by 
breaking the $\mathrm{O}(4)$ symmetry of the NLSM given above down to 
$\mathrm{O}(3)\times Z_{2}$, and studying the effective dynamics of the remnant 
$\mathrm{O}(3)$ NLSM theory in 2+1D.  
Senthil and Fisher~\cite{senthilfisher} argued, however, that our expectations could be belied 
for a special value of the topological angle $2\pi\Theta = \pi$. Here, the ordered phase 
of the $\mathrm{O}(3)$ theory with a unique ground state can be rendered critical due to 
Berry phase-carrying meron-hedgehog and hedgehog instanton tunneling events (see 
Fig.(\ref{hedgehogs})) that interpolate between the ordered 
ground state and an excited state that lies vanishingly close. Remarkably, the single 
meron-hedgehog instanton excitation is unable to lead to the formation of a gap between the 
ground state and the lowest excited state due to a destructive interference mechanism 
arising from the Aharanov-Bohm (AB) like Berry phases (given by the topological angle 
$2\pi\Theta = 2\pi(\Phi_{\mathrm{AB}}/\Phi_{0})=\pi$) accumulated along the multiply 
connected field-configuration manifold that connects these two states. A similar gaplessness of 
the 1+1D $O(3)$ NLSM at $\Theta=\pi$ was also noted earlier~\cite{readshankar}. However, the doubled 
hedgehog instantons undergo a constructive interference by the same token. If the tunneling of 
such doubled hedgehog instantons is RG irrelevant, the ordered $\mathrm{O}(3)$ NLSM theory at 
$2\pi\Theta = \pi$ corresponds to the emergent spin Hall theory in the original electronic 
problem. This is shown in Fig.(\ref{diracscatt}). 

On the other hand, if these doubled instantons 
are RG relevant, they may destabilise the ground state even in the absence of any explicit 
symmetry-breaking perturbations. The RG flow then leads away from the critical theory, opening 
a gap above a doubly 
degenerate ground state of the infra-red (IR) strong coupling fixed point theory. In this case, 
Senthil and Fisher propose that the 
emergent state has no order in the $\mathrm{O}(3)$ order parameter but possesses $Z_{2}$ 
topological order. For an $\mathrm{O}(3)$ order parameter which characterises 
Ne$\acute{e}$l ordering of S$=1/2$ spins in a 2D rectangular lattice, this topological 
state is a $Z_{2}$ spin liquid~\cite{senthilfisher}. In this way, we expect that the emergent spin 
Hall phase observed at criticality in our electronic problem can have an instability leading 
to a state with $Z_{2}$ topological order and a doubly degenerate ground state. Further, this 
state is expected to have fractionalised excitations that make the transitions between the two 
topologically degenerate ground states. Applying explicit symmetry-breaking terms 
to this $\mathrm{O}(3)$ NLSM theory (e.g., through a uniform external B-field) will shift the value 
of the topological angle away from $2\pi\Theta = \pi$. The single instanton excitation will then no 
longer vanish from the interference mechanism. Instead, given that these single instantons are 
typically RG relevant, they will gap the Dirac-spectrum of the electronic problem, leading to the 
two gapped anomalous Hall effect (AHE) theories on either side of the phase transition.

As we will now show, this transition can also be captured by developing a theory 
for the boundary or edge states of this system. The quasiparticle band diagram reveals that the 
phase transition involves a ground 
state level crossing without the bulk gap vanishing everywhere. This is an example of a first 
order phase transition. The presence of the bulk gap allows us to focus on a way to view the 
passage through the band-crossing in the bulk via a theory of the massless 1D Dirac 
electrons at edge~\cite{xuludwig}; this is the essence of the bulk-boundary 
correspondence~\cite{fradkintext}. Thus, we seek the edge theory for when the bulk is at criticality, 
i.e., the case of the anisotropic 2+1D $\mathrm{O}(4)$ NLSM with $\mathrm{O}(3)$ order 
at a value of the topological angle $2\pi\Theta=\pi$. Here, we are guided by the 
fact that the topological term $Q$ of the $\mathrm{O}(4)$ NLSM theory in 2+1D given above 
with $2\pi\Theta=\pi$ is identical to the Wess-Zumino-Novikov-Witten (WZNW) term of the 
1+1D Tomonaga Luttinger liquid for spinless fermions or the Heisenberg spin-1/2 
chain with nearest neighbour antiferromagnetic exchange interactions~\cite{senthilfisher}:
\begin{equation}
\Gamma [U]= \frac{i}{12\pi}\epsilon_{\mu\nu\rho}
\mathrm{tr}[(U^{\dagger}\partial_{\mu}U)(U^{\dagger}\partial_{\nu}U)(U^{\dagger}\partial_{\rho}U)]~,
\end{equation} 
where the field $U$ is an element of the $SU(2)$ group once again, and defines a map from $S^{2}$ 
to $S^{3}$. The topological WZNW term $\Gamma$ is defined in by the area traced out by the field 
$U$ which encloses the volume $S^{3}$. The field $U$ describes a ``superspin" which combines 
the valence bond solid (VBS) order parameter ($\phi_{0}$) and the three component Ne$\acute{e}$l 
order parameter $\vec{\phi}=(\phi_{1},\phi_{2},\phi_{3})$:
\begin{equation}
U= \phi_{0} + i\phi_{1}\sigma_{x} + i\phi_{2}\sigma_{y} + i\phi_{3}\sigma_{z}~~,~~\phi_{0}^{2}+\vec{\phi}^{2}=1~.
\end{equation}
\begin{figure*}[!ht]
\begin{center}
\includegraphics[width=170mm]{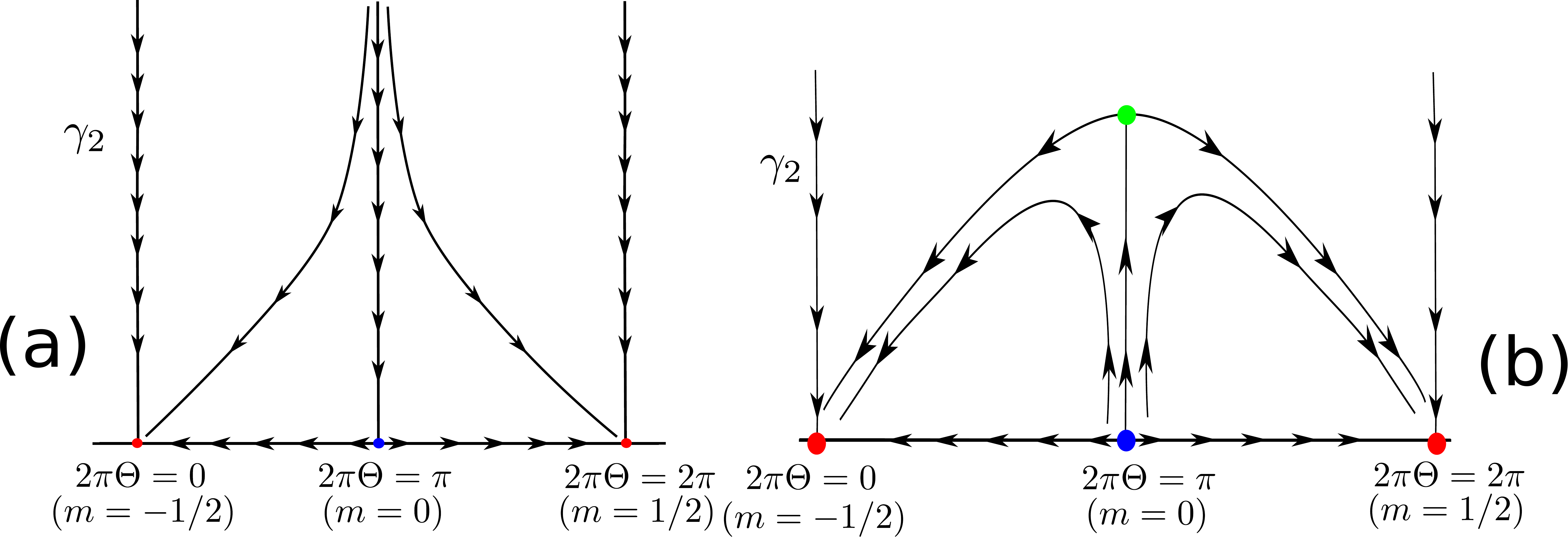}
\caption{(Color  online) The RG phase diagrams obtained from the scaling equations for $\gamma_{2}$ and 
$m$. Recall that the topological angle $\Theta = S - m$, and its scaling can be equivalently seen to track 
that of the single instanton tunnel amplitude $\gamma_{1}$ (see main text for discussion). (a) The RG phase 
diagram for the case of $\gamma_{2}>0$ being marginally irrelevant. The RG flow for $2\pi\Theta=\pi$ leads 
towards the stable weak coupling fixed point theory (blue dot) which corresponds to the emergent spin Hall 
phase in our problem. All RG flows for $2\pi\Theta\neq\pi$ involve an explicit breaking of the emergent 
$SU(2)$ symmetry at $2\pi\Theta=\pi$ via the scaling of the single instanton tunnel amplitude $\gamma_{1}$, 
and lead to the AHE fixed point theories (red dots) with gapped 
ground states at $2\pi\Theta=0,2\pi$ ($m=-1/2,1/2$). Note the symmetry $m\rightarrow -m$ of the RG 
phase diagram about $m=0$. 
$\gamma_{2}$ is RG irrelevant at $2\pi\Theta=0,2\pi$. (b) The RG phase diagram for the case of 
$\gamma_{2}<0$ being marginally irrelevant. The RG flow for $2\pi\Theta=\pi$ leads 
towards the strong coupling fixed point theory (green dot) which corresponds to the $Z_{2}$ topological 
insulator with a doubly degenerate ground state. All other RG flows are as in case (a). The dangerous 
irrelevance of $\gamma_{2}$ and its appearance in the theory only at $2\pi\Theta=\pi$ makes the phase 
transition at this value of the topological angle a case of deconfined quantum criticality.}
\label{deconfrg}\vspace{-2em}
\end{center}
\end{figure*}

Then, following the classic works of Witten~\cite{witten}, as well as Knizhnik and 
Zamolodchikov~\cite{knizhnik}, we may identify the edge state theory at the critical point 
as the $SU(2)_{k=1}$ (i.e., the level-1) WZNW theory of the Heisenberg 
spin-1/2 chain with nearest neighbour antiferromagnetic interactions. As guaranteed by 
the Lieb-Schultz-Mattis (LSM) theorem~\cite{lsm}, this system has a unique ground state 
at the SU(2) symmetric Heisenberg critical point corresponding to the 1D algebraic spin 
liquid~\cite{senthilfisher}. The action for the $SU(2)_{k}$ (i.e., the level-$k$) WZNW theory 
is given by 
\begin{equation}
S = \int d\tau dx\frac{1}{2g}\mathrm{tr}[\partial_{\mu}U^{\dagger}\partial_{\mu}U] + k\Gamma [U]~.
\end{equation}
Witten showed that the topological coupling $k$ affects the RG flow of the NLSM coupling $g$ in 
this WZNW theory~\cite{witten}
\begin{equation}
\frac{dg}{dl} = [1-(\frac{k~g}{4\pi})^{2}]~(\frac{g}{4\pi})~.
\end{equation}
This RG equation shows the existence of a non-trivial stable fixed point at $g^{*}=4\pi/k$. 
Put together with the fact that at this value of $g^{*}$, the theory can be written using the 
non-Abelian bosonisation formalism in terms of free bosons which satisfy a $SU(2)_{k}$ 
Kac-Moody current algebra, Witten conjectured that the WZNW theory must have an exact 
fixed point at $g^{*}$. Using conformal field theoretic methods, this was shown to be correct 
by Knizhnik and Zamolodchikov~\cite{knizhnik}. In this way, we identify the $SU(2)_{k=1}$ WZNW 
theory with NLSM coupling $g^{*}=4\pi$ (corresponding to the topological angle 
$2\pi\Theta=\pi$) as the spin Hall critical theory (with two oppositely directed 1D helical edge 
modes corresponding to massless 1D Dirac electrons of both helicities) at the phase transition 
between the two AHE ordered ground states (i.e., with a gapped bulk and only one chiral 1D 
edge mode of massless Dirac electrons) of our original electronic problem.

But under what conditions is this critical theory stable, and what are its instabilities? As 
shown in Ref.[\cite{fradkintext}], the $SU(2)_{k=1}$ WZNW theory is stable against 
perturbations involving the backscattering coupling between chiral spin currents, 
$g_{1}~\vec{J}_{R}\cdot\vec{J}_{L}$. Here, $\vec{J}_{R/L}$ correspond to the right- and 
left-moving chiral spin currents respectively of the 1+1 edge theory: 
$J^{a}_{R}(x)=\frac{1}{2}\psi^{\dagger}_{R,\sigma}(x)\tau^{a}_{\sigma,\sigma^{'}}\psi_{R,\sigma^{'}}(x)$~,~
$J^{a}_{L}(x)=\frac{1}{2}\psi^{\dagger}_{L,\sigma}(x)\tau^{a}_{\sigma,\sigma^{'}}\psi_{L,\sigma^{'}}(x)$~, 
where $a=(1,2,3)$ and $\tau^{a}$ are the three Pauli matrices. 
The coupling $g_{1}$ corresponds to the amplitude of spin-flip backscattering of electrons between 
the two oppositely directed 1D helical edge states on a given edge of the spin Hall system arising from 
electronic correlations. The RG equation for $g_{1}$ is
\begin{equation}
\frac{dg_{1}}{dl} = -\frac{2}{\pi}g_{1}^{2}~.
\end{equation}
Thus, we can see that the case of a symmetry-preserving perturbation corresponds to 
the coupling $g_{1}>0$: here, $g_{1}$ is marginally irrelevant and can neither break the $SU(2)$ 
symmetry dynamically, nor open a gap in the spectrum. This corresponds to the case of the emergent 
time-reversal symmetry (TRS) associated with the gapless Dirac spectra at the spin Hall critical point 
being robust against perturbations that may break it spontaneously. Conversely, for the case of 
$g_{1}<0$, the $SU(2)$ symmetry of the WZNW theory (i.e., the TRS of the underlying Dirac theory)
is spontaneously broken: $g_{1}$ is now marginally relevant, flows to strong coupling and opens a 
gap in the spectrum above a doubly-degenerate ground state characterised by the spontaneous 
dimerisation of the spins (i.e., valence bond order). This change in RG behaviour with the change in 
the sign of $g_{1}$ is known as the WZNW-type transition~\cite{nakamuravoit}, and needs the 
inclusion of a competing next nearest neighbour antiferromagnetic exchange coupling between the 
spins which exceeds a critical value~(see Ref.(\cite{giamarchi}) and references therein). In the original 
electronic problem, this would correspond to the possibility of gapping the Dirac electrons at the edge 
(i.e., localising them) via scattering from repulsive interactions with extended range, as shown in 
Fig.(\ref{edgescatt}). This can happen 
through Umklapp (i.e., 2-particle) backscattering effects between the 2 helical edge states. Localisation 
via Umklapp scattering mediated by quenched disorder at the edge can also destabilise this 
theory and localise the electrons at the edge~\cite{xumoore,wubernevigzhang}. In this way, we see 
that the transitions seen from theories of the bulk as well as that for the edge 
lead to doubly degenerate ground states with $Z_{2}$ topological order. Interestingly, however, 
the critical theory in the bulk possesses $\mathrm{O}(3)$ order in the ground state while that at the 
edge is the fluctuation disordered algebraic spin-liquid state of the $SU(2)_{1}$ WZNW theory. 
This difference is guaranteed by the Mermin-Wagner-Hohenberg theorem~\cite{merminwagnerhohenberg} which forbids ordering 
at any finite temperature in 2D systems, in conjunction with 
the Lieb-Schultz-Mattis (LSM)~\cite{lsm} theorem for the 1+1D Heisenberg spin-1/2 chain 
with nearest neighbour antiferromagnetic exchange interactions. 
\begin{figure*}[!ht]
\begin{center}
\includegraphics[width=170mm]{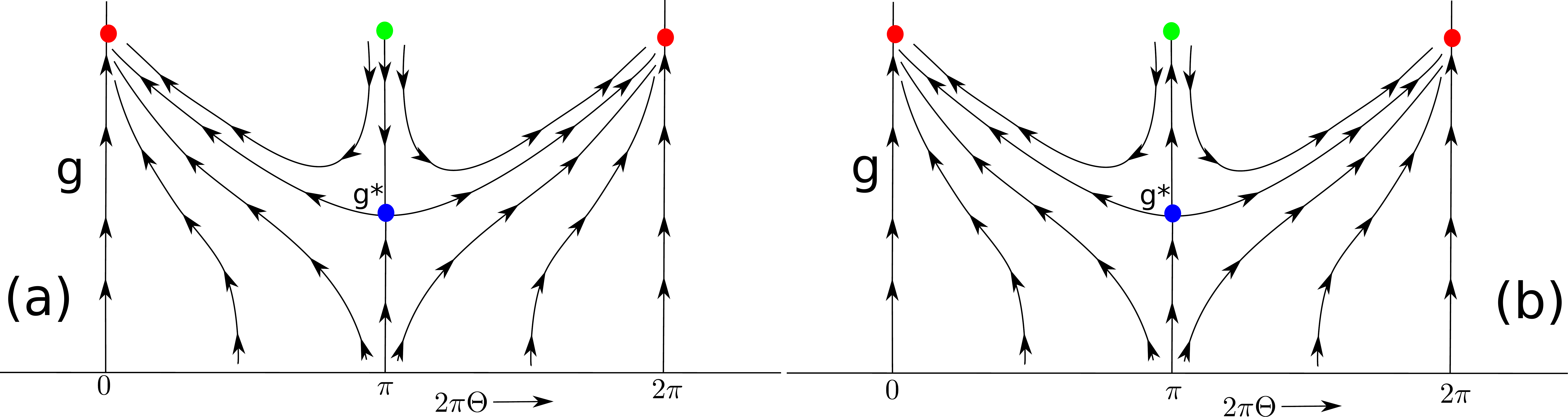}
\caption{(Color  online) The RG phase diagrams obtained from the scaling equations for the NLSM coupling $g$ and 
the topological angle $\Theta$. (a) The RG phase diagram for the case of the doubled instanton tunnel coupling 
$\gamma_{2}>0$ being marginally irrelevant. The RG flows for $2\pi\Theta=\pi$ leads towards the intermediate 
coupling fixed point theory at $g=g^{*}=4\pi$ (blue dot) 
which corresponds to the emergent spin Hall phase in our problem. All RG flows for $2\pi\Theta\neq\pi$ lead to the 
AHE fixed point theories (red dots) with gapped ground states at $2\pi\Theta=0,2\pi$. Note the symmetry of the RG 
phase diagram $\Theta\rightarrow -\Theta$ about $\Theta=1/2$. $g$ is RG relevant at $2\pi\Theta=0,2\pi$. (b) The 
RG phase diagram for the case of $\gamma_{2}<0$ being marginally irrelevant. The RG flow for $2\pi\Theta=\pi$ now 
leads away from the intermediate fixed point at $g^{*}$ towards the strong coupling fixed point theory (green dot) 
which corresponds to the $Z_{2}$ topological insulator with a doubly degenerate ground state. All other RG flows are 
as in case (a). The change in the nature of the RG flow about $g^{*}$ for $2\pi\Theta=\pi$ brought about by the 
dangerous irrelevance of $\gamma_{2}$ makes the phase 
transition at this value of the topological angle a case of deconfined quantum criticality.}
\label{tqptrg}\vspace{-2em}
\end{center}
\end{figure*}

We can now address the perturbations that appear upon breaking the global $SU(2)$ symmetry 
of the action, i.e., by departing from the special value of the topological angle $2\pi\Theta=\pi$. 
This will help in displaying the fact that there exists an analog of the subtle interference 
mechanism discussed earlier for instantons of the 2+1D $\mathrm{O}(3)$ NLSM which also 
renders the 1+1D WZNW theory with $2\pi\Theta=\pi$ stable. Following Affleck~\cite{affleck}, 
it is convenient to consider the role played by topological excitations through the sine-Gordon 
version of the 1+1D $\mathrm{O}(3)$ NLSM Lagrangian:
\begin{equation}
L = \frac{1}{2}(\partial_{\mu}\phi)^{2} + \sum_{n, q_{n}}\gamma_{n}\exp^{i(n\sqrt{g}\phi + 2\pi n q_{n}\Theta)} - \frac{m}{\pi}\partial_{x}\phi~.
\end{equation}
Here, $\phi$ is the scalar field encoding the Ne$\mathrm{\grave{e}}$l order, $g$ is the phase 
stiffness parameter/NLSM coupling, $n$ is the vorticity of the topological excitations in the field $\phi$, $q_{n}$ 
is the charge of the topological excitation and $\gamma_{n}$ the fugacity for an 
instanton excitation with vorticity $n$, and the topological angle is $\Theta = S - m$~\cite{tanakatotsukahu}. 
Note that $S$ is the spin quantum number of the constituent spins in the problem and $m$ is 
the magnetisation; $S=1/2$ corresponds to the case of gapless Dirac electrons and $m\equiv m_{z}-m_{z}^{*}$ 
in the original electronic problem. The first term gives the cost of generating collective excitations, 
the second and third the cost of topological excitations and the fourth the effect of an explicit 
symmetry-breaking B-field (through the magnetisation $m$). 
The bare value of the instanton fugacity can be computed using standard instanton techniques~\cite{rajaraman}, 
and $\gamma_{1}\sim\exp(-S_{0}/\hbar)$, where $S_{0}$ is the classical Euclidean action for the 
instanton of the sine-Gordon problem. As shown by 
Affleck~\cite{affleck}, this simplest instanton excitation has $n=\pm 1$ and each value of $n$ 
has two charges $q_{n}=\pm 1/2$ (which corresponds to the two values of the Chern no. $\tilde{N}_{3}$ 
discussed earlier for the electronic problem). In the absence of an external B-field, $m=0$, and with $S=1/2$, 
we have $\Theta=1/2$. Then, in this case, we have
\begin{eqnarray}
&&\sum_{n=\pm 1, q_{n}=\pm 1/2}\gamma_{1}\exp^{i(n\sqrt{g}\phi + 2\pi n q_{n}\Theta)} \nonumber\\
&=&\gamma_{1}(\exp^{i\pi/2}+\exp^{-i\pi/2})(\exp^{i\sqrt{g}\phi} +\exp^{-i\sqrt{g}\phi})\nonumber\\
&=&4\gamma_{1}\cos(\pi/2)\cos(\sqrt{g}\phi)~=~0~.
\end{eqnarray}
In this way, we can see that while these instanton excitations are RG relevant (the coupling $\gamma_{1}$ 
has a scaling dimension $g<1/2$), it is suppressed via a destructive interference mechanism for $2\pi\Theta=\pi$ 
and thus unable to open a gap in the spectrum~\cite{affleck} (consult Fig.(\ref{diracscatt})(a)). However, the 
instantons with doubled vorticity 
$n=\pm 2$ with topological charges $q=\pm 1$ are hedgehog excitations (where $|q|=1$ corresponds to 
the value of the Chern no. $N_{3}$ discussed earlier for the case of massless Dirac electrons) which undergo 
a constructive interference even at $2\pi\Theta=\pi$ by the same token (consult Fig.(\ref{diracscatt})(b))
\begin{eqnarray}
&&\sum_{n=\pm 2, q_{n}=\pm 1}\gamma_{2}\exp^{i(n\sqrt{g}\phi + 2\pi n q_{n}\Theta)} \nonumber\\
&=&4\gamma_{2}(\exp^{i2\pi}+\exp^{-i2\pi})(\exp^{i2\sqrt{g}\phi} +\exp^{-i2\sqrt{g}\phi})\nonumber\\
&=&4\gamma_{2}\cos(2\pi)\cos(2\sqrt{g}\phi)~=~4\gamma_{2}\cos(2\sqrt{g}\phi)~.
\end{eqnarray}
The RG equations for $g$ and $\gamma_{2}$ are given by 
\begin{equation}
\frac{dg}{dl} = -\frac{\gamma_{2}^{2}}{4}~~,~~\frac{d\gamma_{2}}{dl} = (2-4g)\gamma_{2}~.
\end{equation}
As discussed earlier, there is no RG equation for the $\Theta$ parameter for the case of $m=0$ due to the 
existence of an extra discrete symmetry of the theory. While the two RG equations given above describe
the Berezinskii-Kosterlitz-Thouless (BKT)~\cite{berezinskiikosterlitzthouless} universality class of transitions, it 
also describes the WZNW universality class. This can be seen as follows~\cite{nakamuravoit}: first denote 
$y_{0} = 4g-2$~,~$y_{1}=\gamma_{2}$, then impose the global $SU(2)$ symmetry of 
the WZNW theory on the sine-Gordon problem by requiring that $y_{0}=y_{1}\equiv y$ even under the RG 
transformations. This then gives the RG equation for the lone coupling in the problem as $dy/dl = -y^{2}$, 
which is precisely the RG observed for the coupling $g_{1}$ of the WZNW theory seen earlier with the 
redefinition: $g_{1}=(\pi/2)y$. From our earlier discussion for $g_{1}$, we can see that the coupling $y$ is 
either marginally irrelevant or marginally relevant, depending on its sign. Further, the $SU(2)$ symmetry 
of the WZNW fixed point fixes the value of the coupling $g$ of the sine-Gordon theory as 
$g^{*}=1/2$~\cite{giamarchi}: this is where the coupling $\gamma_{2}$ is exactly marginal. The RG phase 
diagrams for the BKT and WZNW transitions is shown in Fig.(\ref{bktwznwrg}).

From the identification of the topological angle $\Theta=S-m$~\cite{tanakatotsukahu}, we can also perform 
a similar RG analysis of the sine-Gordon model for $m\neq 0$. First, we can see immediately that the 
destructive interference mechanism no longer suppresses the instanton tunnel coupling 
$\gamma_{1}$ for $m\neq 0$:
\begin{eqnarray}
&&\sum_{n=\pm 1, q_{n}=\pm 1/2}\gamma_{1}\exp^{i(n\sqrt{g}\phi + 2\pi n q_{n}\Theta)} \nonumber\\
&=&\gamma_{1}(\exp^{i\pi(1/2-m)}+\exp^{-i\pi(1/2-m)})(\exp^{i\sqrt{g}\phi} +\exp^{-i\sqrt{g}\phi})\nonumber\\
&=&4\gamma_{1}\cos(\pi(1/2-m))\cos(\sqrt{g}\phi)~.
\end{eqnarray}
The RG equation for $g$, $\gamma_{1}$ and the magnetisation $m$ is given as~\cite{horowitz,giamarchi}
\begin{eqnarray}
\frac{dg}{dl} &=& -\gamma_{1}^{2} J_{0}(m\alpha)~~,~~\frac{d\gamma_{1}}{dl} = (2-g)\gamma_{1}\nonumber\\
\frac{dm}{dl} &=& m - \frac{\gamma_{1}^{2}}{2\pi\alpha} J_{1}(m\alpha)~,
\end{eqnarray}
where $J_{0}$ and $J_{1}$ are Bessel functions that arise from the use of sharp cut-off functions while implementing 
the RG transformations~\cite{giamarchi}. Importantly, the RG equation for $m$ is symmetric under the interchange 
of $m\rightarrow -m$. These RG equations show that the presence of a non-zero $m$ leads to two gapped ground 
states in which the $SU(2)$ symmetry is explicitly broken and the magnetisation is saturated at $m=-1/2$ and $m=1/2$ 
respectively. A large magnetisation will, in turn, lead to a suppression the instanton tunnel amplitude $\gamma_{1}$. This 
is easily seen in the original electronic problem, where a growing magnetisation $m$ of the spin problem corresponds to 
the presence of only those massless Dirac electrons at the edge whose spins are aligned with the direction of $m_{z}$. 
Backscattering of these electrons (with an amplitude corresponding to $\gamma_{1}$) is strongly suppressed due to 
the conservation of their helicity, and is common to other cases in which spin rotation symmetry is explicitly broken 
by a strong external magnetic field: the chiral massless Dirac electrons observed at the edge of the quantum Hall 
effect~\cite{wen} and electrons in 1D Tomonaga-Luttinger liquid quantum wire with a large Zeeman gap~\cite{lalsen}. 
The RG phase diagram for the doubled instanton tunnel amplitude $\gamma_{2}$ versus the topological angle $\Theta$ 
is shown in Fig.(\ref{deconfrg}). 

Note that the values of $m=-1/2,1/2$ correspond to values of the topological angle $\Theta=0,1$. That 
the theories at $\Theta=0,1$ possess gapped spectra is 
corroborated by the works of Polyakov~\cite{polyakov} and Haldane~\cite{haldane}, who showed that the 1+1D 
$\mathrm{O}(3)$ NLSM theory with a value of the topological angle $2\pi\Theta=0,~2\pi$ has the NLSM coupling $g$ 
flow to strong coupling under RG. It is then straightforward to identify these two strong coupling fixed points at 
$\Theta=0,1$ as the two AHE theories (with bulk gaps, and a single edge mode of chiral massless Dirac electrons) on 
either side of the phase transition in our original electronic problem, $m_{z}<<m_{z}^{*}$ and $m_{z}>>m_{z}^{*}$. 
In this way, we are able to see the manner in which instanton excitations drive the Lifshitz transition, changing the
topology of the electronic momentum space.

Finally, similar considerations can also be 
made for the effects of a staggered magnetic field on the system: as discussed by Affleck~\cite{afflecknpb}, such a field 
couples to the staggered magnetisation $m_{N}$ and will again change the $\Theta$ topological angle, breaking the 
global $SU(2)$ symmetry and involve a doubling of the unit cell via the establishment of Ne$\acute{e}$l order. Further, 
it can be shown~\cite{senlal} that in the presence of explicit dimerisation (i.e. dimer order) in the system, the doubled 
instanton tunnel amplitude $\gamma_{2}$ will affect the RG equation for the staggered magnetisation $m_{N}$ in a 
way analogous to that shown above for the way in which $\gamma_{1}$ affects the RG of the uniform magnetisation 
$m$. We show the RG phase diagram for the NLSM coupling $g$ versus the topological angle $\Theta$ in 
Fig.(\ref{tqptrg}). Importantly, via the bulk-boundary correspondence that is well established for topological 
insulators~\cite{fradkintext}, our findings from the edge field theories give evidence for the fact that the dangerous 
irrelevance of the doubled instanton tunnel amplitude $\gamma_{2}$ observed only at $2\pi\Theta=\pi$, and the 
subsequent stabilisation of the spin Hall phase in the bulk, is a case of deconfined quantum 
criticality~\cite{senthilfisher,fradkintext}. The spin Hall theory for $\gamma_{2}=0$ is a critical fixed point, with three 
possible unstable directions in terms of the NLSM coupling $g$, the topological angle $\Theta$ and the fugacity for the 
doubled instantons $\gamma_{2}$. We can, therefore, classify the spin Hall phase as tricritical~\cite{goldenfeld}, with the 
transition being first order (i.e., involving explicit symmetry breaking) in $\Theta$ and continuous in $g$ as well as $\gamma_{2}$. 
  
\subsection{Self-consistent BdG formalism in momentum space}
Hamiltonian (1) in the main  text is diagonalized via a spin-generalized
Bogoliubov-Valatin transformation
$\hat{c}_{k\sigma}=\sum_{n\sigma^{\prime}}u_{n\sigma\sigma^{\prime}}
(\mathbf{k})\hat{\gamma}_{n\sigma^{\prime}}+
v_{n\sigma\sigma^{\prime}}^*(\mathbf{k})\hat{\gamma}_{n\sigma^{\prime}}^{\dagger}$,
where           $u_{n\sigma\sigma^{\prime}}(\mathbf{k})$           and
$v_{n\sigma\sigma^{\prime}}(\mathbf{k})$    are   quasi-particle   and
quasi-hole            amplitudes            respectively           and
$\hat{\gamma}_{n\sigma^{\prime}}$ is a fermionic operator.

The              mean-field              pairing             amplitude
$\Delta=-U<c_{k\uparrow}c_{-k\downarrow}>$, where  $U$ is the pairwise
attractive  interaction  potential,  is  obtained via  the  Bogoliubov
amplitudes   $u_{n\sigma}(\mathbf{k})$  and  $v_{n\sigma}(\mathbf{k})$
($\sigma^{\prime}$, being  a pseudo-index, is  omitted for simplicity)
as
\begin{equation}
\begin{split}
\Delta(\mathbf{k})&=-U\sum_n[u_{n\uparrow}(\mathbf{k})v_{n\downarrow}^*(\mathbf{-k})
(1-f(E_n))\\
&+u_{n\downarrow}(\mathbf{k})v_{n\uparrow}^*(\mathbf{-k})f(E_n)]
\end{split}
\end{equation}
where $f(x)=1/(1+e^{x/(k_BT)})$ is the Fermi function at temperature 
$T$ and $k_B$ is the Boltzmann constant.\vspace{2em}

\subsection{Self-consistent BdG formalism in real space}
The  conventional route to studying  disordered superconductivity  is to
solve the self-consistent BdG equations
${\cal H}\phi_n(r_i)=\epsilon_n\phi_n(r_i)$
where
$\phi_n=[u_{n\uparrow}(r_i),u_{n\downarrow}(r_i),v_{n\uparrow}(r_i),v_{n\downarrow}(r_i)]$
at every site using Bogoliubov transformation
$\hat{c}_{i\sigma}(r_i)=\sum_{i,\sigma^{\prime}}u_{n\sigma\sigma^{\prime}}(r_i)\hat{\gamma}_{n\sigma^{\prime}}+v_{n\sigma\sigma^{\prime}}^*(r_i)\hat{\gamma}^{\dagger}_{n\sigma^{\prime}}$.   
As above, the mean field pairing amplitude
$\Delta(r_i)=-U<c_{i\uparrow}c_{i\downarrow}>$ and magnetization
$m(r_i)=<c_{i\uparrow}^{\dagger}c_{i\uparrow}-c_{i\downarrow}^{\dagger}c_{i\downarrow}>$
are calculated from the following relations:
\begin{equation}
\begin{split}
\Delta(r_i)=&-U\sum_{n}[u_{n\uparrow}(r_i)v^*_{n\downarrow}(r_i)(1-f(E_n))\\
&+u_{n\downarrow}(r_i)v^*_{n\uparrow}(r_i)f(E_n)]
\end{split}
\label{delsi}
\end{equation}\vspace{-1em} 
\begin{equation}
m(r_i)=\sum_{n,\sigma}u_{n\uparrow}^{*}u_{n\downarrow}f(E_n)+v_{n\uparrow}v_{n\downarrow}^{*}(1-f(E_n))
\end{equation}
where $f(x)=1/(1+e^{x/(k_BT)})$ is the Fermi function at temperature 
$T$ and $k_B$ is the Boltzmann constant.

\subsection{Hamiltonian for Magnetic Impurity}
The  interaction  between  the  magnetic impurity  and  the  itinerant
electrons can be represented by the Hund's coupling-like term
$\hat{H}_{mag}=-J_H\sum_{j}\vec{S}_j\cdot\vec{s}_j$, where  $\vec{S}_j$ is the impurity
spin     and      $\vec{s}_j=\frac{1}{2}c_{j\sigma}^{\dagger}\vec{\sigma}_{\sigma
  \sigma^{\prime}}c_{j\sigma^{\prime}}$   represents   the   intrinsic
electron spin.  Assuming the impurity is represented by a 
classical spin of unit amplitude and polar angles    
$\theta$ and $\phi$, the mean-field $\hat{H}_{mag}$ is 
\begin{equation}\vspace{-1em}
\hat{H}_{mag}=-\frac{J_H}{2}\begin{pmatrix} \begin{array}{cc}
    c_{j\uparrow}^{\dagger} &
    c_{j\downarrow}^{\dagger} \end{array} \end{pmatrix} 
\Pi
\begin{pmatrix} \begin{array}{cc}
    c_{j\uparrow} \\
    c_{j\downarrow} \end{array} \end{pmatrix}
\end{equation}\vspace{0em}
\noindent{where},
\begin{equation*}
\Pi=\begin{pmatrix} \begin{array}{cc}
    \cos{\theta} & \sin{\theta}\cos{\phi}-i\sin{\theta}\sin{\phi} \\
    \sin{\theta}\cos{\phi}+i\sin{\theta}\sin{\phi} &
    -\cos{\theta} \end{array} \end{pmatrix}
\end{equation*}
If we  assume that the impurity  spin is aligned  perpendicular to the
2D-plane (\textit{i.e.}, $\theta=0$), we obtain
$\hat{H}_{mag}=-J_{H}/2\sum_{j,\sigma,\sigma^{\prime}}       \boldsymbol{\sigma}_{\sigma
  \sigma^{\prime}}^{z}      c_{j\sigma}^{\dagger}c_{j\sigma^{\prime}}$,
where $j$ runs  over $N_d$ number of impurity  sites, randomly located
in   the   two-dimensional  space.


\begin{thebibliography}{10}

\bibitem{PhysRevLett.92.126603}
Sinova,~J. {\em et~al.}
\newblock Universal Intrinsic Spin Hall Effect. {\it Phys. Rev. Lett.} {\bf 92}, 126603 (2004).

\bibitem{RevModPhys.82.1539}
Nagaosa,~N., Sinova,~J., Onoda,~S., MacDonald,~A.~H., and Ong,~N.~P.,
\newblock Anomalous Hall effect. {\it Rev. Mod. Phys.} {\bf 82}, 1539 (2010).

\bibitem{PhysRevB.77.220501}
Fujimoto,~S.
\newblock Topological order and non-Abelian statistics in noncentrosymmetric $s$-wave superconductors. 
{\it Phys. Rev. B} {\bf 77}, 220501 (2008).

\bibitem{PhysRevLett.101.160401}
Zhang,~C., Tewari,~S., Lutchyn,~R.~M. and Das~Sarma,~S.
\newblock ${p}_{x}+i{p}_{y}$ Superfluid from $s$-Wave Interactions of Fermionic Cold Atoms. {\it Phys. Rev. Lett.} {\bf 101}, 160401 (2008).

\bibitem{Richter2013}
Richter,~C. {\em et~al.}
\newblock Interface superconductor with gap behaviour like a high-temperature superconductor. {\it Nature} {\bf 502}, 528 (2013),
\newblock Letter.

\bibitem{Caviglia2008}
Caviglia,~A.~D. {\em et~al.}
\newblock Electric field control of the {L}a{A}l{O}$_3$/{S}r{T}i{O}$_3$ interface ground state. {\it Nature} {\bf 456}, 624 (2008).

\bibitem{PhysRevB.53.4393}
Allen,~P.~B. {\em et~al.}
\newblock Transport properties, thermodynamic properties, and electronic structure of SrRuO. {\it Phys. Rev. B} {\bf 53}, 4393 (1996).

\bibitem{PhysRevB.70.180407}
Kats,~Y.,~Genish,~ I.~,Klein,~ L.~Reiner,~ J.~W. and Beasley,~M.~R.
\newblock Testing the Berry phase model for extraordinary Hall effect in SrRuO. {\it Phys. Rev. B} {\bf 70}, 180407 (2004).

\bibitem{JPSJ.66.3893}
Izumi,~M.~,Nakazawa,~ K.~, Bando,~Y.~,Yoneda,~ Y.~ and Terauchi,~H.
\newblock Magnetotransport of $SrRuO_{3}$ Thin Film on $SrTiO_{3}~(001)$. {\it J. Phys. Soc. Jpn.} {\bf 66}, 3893 (1997).

\bibitem{PhysRevLett.93.016602}
Mathieu,~R.~ {\em et~al.}
\newblock Scaling of the Anomalous Hall Effect in $Sr_{1-x}Ca_{x}RuO_{3}$. {\it Phys. Rev. Lett.} {\bf 93}, 016602 (2004).

\bibitem{PhysRevLett.88.207208}
Jungwirth,~T.~, Niu,~Q.~ and MacDonald,~ A.~H.
\newblock Anomalous Hall Effect in Ferromagnetic Semiconductors. {\it Phys. Rev. Lett.} {\bf 88}, 207208 (2002).

\bibitem{JPSJ.70.2999}
Oda,~K.~ {\em et~al.}
\newblock Unusual Anomalous Hall Resistivities of $CuCr_{2}S_{4}$, $Cu_{0.5}Zn_{0.5}Cr_{2}Se_{4}$ and $Cr_{3}Te_{4}$. 
{\it J. Phys. Soc. Jpn.} {\bf 70}, 2999 (2001).

\bibitem{Lee12032004}
Lee,~W.-L.~, Watauchi,~S.~, Miller,~ V.~L., Cava,~R.~J. and Ong,~N.~P.
\newblock Dissipationless Anomalous Hall Current in the Ferromagnetic Spinel $CuCr_{2}Se_{4-x}Br_{x}$. 
{\it Science} {\bf 303}, 1647 (2004).

\bibitem{PhysRevB.66.174429}
Galanakis,~I.~, Dederichs,~ P.~H. and Papanikolaou,~N.
\newblock Slater-Pauling behaviour and origin of the half-metallicity of the full-Heusler alloys. {\it Phys. Rev. B} {\bf 66}, 174429 (2002).

\bibitem{PhysRevB.70.205114}
Block,~T.~, Carey,~M.~J., Gurney,~B.~A. and Jepsen,~O.
\newblock Band-structure calculations of the half-metallic ferromagnetism and structural stability of full- and half-Heusler phases. 
{\it Phys. Rev. B} {\bf 70}, 205114 (2004).

\bibitem{PhysRevB.77.014433}
Checkelsky,~J.~G.~, Lee,~ M.~, Morosan,~ E.~, Cava,~R.~J. and Ong,~N.~P.
\newblock Anomalous Hall effect and magnetoresistance in the layered ferromagnet $Fe_{14}TaS_{2}$: The inelastic regime. 
{\it Phys. Rev. B} {\bf 77}, 014433 (2008).

\bibitem{PhysRev.95.1154}
Karplus,~R.~ and Luttinger,~J.~M. 
\newblock Hall Effect in Ferromagnetics. {\it Phys. Rev.} {\bf 95}, 1154 (1954).

\bibitem{Fang03102003}
Fang,~Z.~ {\em et~al.}
\newblock The Anomalous Hall Effect and Magnetic Monopoles in Momentum Space. {\it Science} {\bf 302}, 92 (2003).

\bibitem{OnodaJPSJ.71.19}
Onoda,~M.~ and Nagaosa,~N. 
\newblock Topological Nature of Anomalous Hall Effect in Ferromagnets. {\it J. Phys. Soc. Jpn.} {\bf 71}, 19  (2002).

\bibitem{PhysRevB.2.4559}
Berger,~L.
\newblock Side-Jump Mechanism for the Hall Effect of Ferromagnets. {\it Phys. Rev. B} {\bf 2}, 4559 (1970).

\bibitem{Smit1955877}
Smit,~J.
\newblock The spontaneous hall effect in ferromagnetics: I. {\it Physica} {\bf 21}, 877  (1955).

\bibitem{PhysRevLett.97.126602}
Onoda,~S.~, Sugimoto,~N.~ and Nagaosa,~ N.
\newblock Intrinsic Versus Extrinsic Anomalous Hall Effect in Ferromagnets. {\it Phys. Rev. Lett.} {\bf 97}, 126602 (2006).

\bibitem{0022-3719-17-33-015}
Bychkov,~Y.~A. and Rashba,~E.~I.
\newblock Oscillatory effects and the magnetic susceptibility of carriers in inversion layers. 
{\it Journal of Physics C: Solid State Physics} {\bf 17}, 6039 (1984).

\bibitem{PhysRevB.71.224423}
Dugaev,~V.~K.~, Bruno,~ P.~, Taillefumier,~ M.~, Canals,~B.~ and Lacroix,~C.
\newblock Anomalous Hall effect in a two-dimensional electron gas with spin-orbit interaction. 
{\it Phys. Rev. B} {\bf 71}, 224423 (2005).

\bibitem{PhysRevLett.102.086602}
Kontani,~H.~, Goryo,~J.~ and Hirashima,~D.~S.
\newblock Intrinsic Spin Hall Effect in the $s$-Wave Superconducting State: Analysis of the Rashba Model. 
{\it Phys. Rev. Lett.} {\bf 102}, 086602 (2009).

\bibitem{gradhandannett}
Gradhand,~M.~ and Annett,~J.~F.
\newblock The Berry curvature of the Bogoliubov quasiparticle Bloch states in the unconventional superconductor 
$Sr_{2}RuO_{4}$. {\it J. Phys.: Condens. Matter.} {\bf 26}, 274205 (2014). 

\bibitem{chungroy}
Chung,~S.~B.~ and Roy,~R.
\newblock Hall conductivity in the normal and superconducting phases of the Rashba system with Zeeman field.
arXiv:1407.3883

\bibitem{PhysRevB.68.045327}
Culcer,~D.~, MacDonald,~A.~H.~ and Niu,~Q.
\newblock Anomalous Hall effect in paramagnetic two-dimensional systems. {\it Phys. Rev. B} {\bf 68}, 045327 (2003).

\bibitem{lifshitz}
Lifshitz,~I.~M., Anomalies of electron characteristics of a metal in the high pressure region. 
{\it Sov. Phys. JETP} {\bf 11}, 1130-1135 (1960).

\bibitem{PhysRevB.81.125318}
Alicea,~J.
\newblock Majorana fermions in a tunable semiconductor device. {\it Phys. Rev. B} {\bf 81}, 125318 (2010).

\bibitem{volovik}
Volovik,~G.~E.
\newblock {\it The Universe in a Helium Droplet} (Oxford Univ. Press, Oxford, 2003).

\bibitem{Anderson195926}
Anderson,~P.~W.
\newblock Theory of dirty superconductors. {\it Journal of Physics and Chemistry of Solids} {\bf 11}, 26  (1959).

\bibitem{dubinature2007}
Dubi,~Y.~, Meir,~Y.~ and Avishai,~Y.
\newblock Nature of the superconductor-insulator transition in disordered superconductors. 
{\it Nature} {\bf 449}, 876 (2007).

\bibitem{ishikawamatsuyama}
Ishikawa,~K.~ and Matsuyama,~T.
\newblock A microscopic theory of the quantum Hall effect. {\it Nuclear Phys. B} {\bf 280}, 523 (1987).

\bibitem{mohanta_phase_segregation}
Mohanta,~N.~ and Taraphder,~A.
\newblock Phase segregation of superconductivity and ferromagnetism at the LaAlO 3 /SrTiO 3 interface. 
{\it J Phys: Condens. Matter} {\bf 26}, 025705 (2014);
\textit{ibid}, {\bf 26}, 215703 (2014).

\bibitem{dagotto}
Dagotto,~E. 
\newblock Complexity in strongly correlated electronic systems. {\it Science} {\bf 309}, 257 (2005).

\bibitem{fernandesschmalian}
Fernandes,~R.~M.~ and Schmalian,~J.
\newblock Complex critical exponents for percolation transitions in Josephson-junction arrays, antiferromagnets, and interacting bosons. 
{\it Phys. Rev. Lett.} {\bf 106}, 067004 (2011).

\bibitem{PhysRevB.83.184520}
Potter,~A.~C.~ and Lee,~P.~A. 
\newblock Engineering a $p+\mathit{ip}$ superconductor: Comparison of topological insulator and Rashba spin-orbit-coupled materials. 
{\it Phys. Rev. B} {\bf 83}, 184520 (2011).

\end{thebibliography}

\begin{thebibliography}{10}

\bibitem{volovik}
Volovik,~G.~E.  
\newblock {\it The Universe in a Helium Droplet} (Oxford Univ. Press, Oxford, 2003).

\bibitem{kramers}
Kramers,~H.~A. 
\newblock Th$\acute{\mathrm{e}}$orie g$\acute{\mathrm{e}}$n$\acute{\mathrm{e}}$rale de la rotation 
paramagn$\acute{\mathrm{e}}$tique dan le cristaux. {\it Proc. Amsterdam Acad.} {\bf 33}, 959 (1930);
Sakurai,~J.~J. 
\newblock {\it Modern Quantum Mechanics} (Addison-Wesley, Reading, MA, 1994).

\bibitem{sinova}
Sinova,~J. {\em et~al.}
\newblock Universal Intrinsic Spin Hall Effect. {\it Phys. Rev. Lett.} {\bf 92}, 126603 (2004).

\bibitem{kanemele}
Kane,~C.~L. ~ and Mele,~E.~J.  
\newblock $Z_{2}$ Topological Order and the Quantum Spin Hall Effect. {\it Phys. Rev. Lett.} {\bf 95}, 146802 (2005).

\bibitem{senthilfisher}
Senthil,~T.~ and Fisher,~M.~P.~A. 
\newblock Competing orders, nonlinear sigma models, and topological terms in quantum magnets. 
{\it Phys. Rev. B} {\bf 74}, 064405 (2006).

\bibitem{abanovwiegmann}
Abanov,~A.~G.~ and Wiegmann,~P.~B.  
\newblock Theta-terms in nonlinear sigma-models. {\it Nucl. Phys. B} {\bf 570}, 685 (2000).

\bibitem{xuludwig}
Xu,~C.~ and Ludwig,~A.~W.~W. 
\newblock Nonperturbative Effects of a Topological Theta Term on Principal Chiral Nonlinear Sigma Models in $2+1$ Dimensions. 
{\it Phys. Rev. Lett.} {\bf 110}, 200405 (2013).

\bibitem{lifshitz}
Lifshitz,~I.~M., Anomalies of electron characteristics of a metal in the high pressure region. 
{\it Sov. Phys. JETP} {\bf 11}, 1130-1135 (1960).

\bibitem{rajaraman}
Rajaraman,~R.
\newblock {\it Solitons and Instantons: An Introduction to Solitons and Instantons in Quantum Field Theory} (North Holland Personal Library, Amsterdam, 1987).

\bibitem{readshankar}
Shankar,~R.~and~Read,~N.
\newblock The $\Theta=\pi$ nonlinear sigma model is massless. {\it Nucl. Phys. B} {\bf 336}, 457 (1990);
Controzzi,~D.~and~Mussardo,~G.
\newblock Mass Spectrum of the Two-Dimensional $O(3)$ Sigma Model with a $\Theta$ Term. {\it Phys. Rev. Lett.} {\bf 92}, 021601 (2004).

\bibitem{fradkintext}
Fradkin,~E. 
\newblock {\it Field Theories of Condensed Matter Physics} (2nd Edition, Cambridge Univ. Press, Cambridge, 2013).

\bibitem{witten}
Witten,~E. 
\newblock Nonabelian Bosonization in Two-Dimensions. {\it Commun. Math. Phys.} {\bf 92}, 455 (1984).

\bibitem{knizhnik}
Knizhnik,~V.~G.~ and Zamolodchikov,~A.~B.  
\newblock Current Algebra and Wess-Zumino Model in Two-Dimensions. {\it Nucl. Phys. B} {\bf 247}, 83 (1984).

\bibitem{lsm} 
Lieb,~E.,~ Schultz,~ T. and Mattis,~D. C.
\newblock Two Soluble Models of an Antiferromagnetic Chain. {\it Ann. Phys. (N.Y.)} {\bf 16}, 407 (1961). 

\bibitem{nakamuravoit}
Nakamura,~M.~ and Voit,~J.
\newblock Lattice twist operators and vertex operators in sine-Gordon theory in one dimension. {\it Phys. Rev. B} {\bf 65}, 153110 (2002)

\bibitem{giamarchi}
Giamarchi,~T. 
\newblock {\it Quantum Physics in One Dimension} (Oxford Univ. Press, Oxford, 2003).

\bibitem{xumoore}
Xu,~C.~ and Moore,~J. E.
\newblock Stability of the quantum spin Hall effect: Effects of interactions, disorder, and $Z_{2}$ topology. {\it Phys. Rev. B} {\bf 73}, 045322 (2006).

\bibitem{wubernevigzhang}
Wu,~C.,~ Bernevig,~A. and Zhang,~S.-C.
\newblock Helical Liquid and the Edge of Quantum Spin Hall Systems. {\it Phys. Rev. Lett.} {\bf 96}, 106401 (2006).

\bibitem{merminwagnerhohenberg}
Mermin,~N.~D.~ and Wagner,~H. 
\newblock Absence of Ferromagnetism or Antiferromagnetism in One- or Two-Dimensional Isotropic Heisenberg Models. 
{\it Phys. Rev. Lett.} {\bf 17}, 1133 (1966);
Hohenberg,~P.~C. 
\newblock Existence of Long-Range Order in One and Two Dimensions. {\it Phys. Rev.} {\bf 158}, 383 (1967).

\bibitem{affleck}
Affleck,~I.  
\newblock Mass Generation by Merons in Quantum Spin Chains and the $O(3)$ $\sigma$ Model. 
{\it Phys. Rev. Lett.} {\bf 56}, 408 (1986); {\it ibid.}, 
\newblock Quantum spin chains and the Haldane gap. {\it J. Phys. Cond. Matt.} {\bf 1}, 3047 (1989).

\bibitem{tanakatotsukahu}
Tanaka,~A.,~Totsuka,~K.~ and Hu,~X.
\newblock Geometric phases and the magnetization process in quantum antiferromagnets. {\it Phys. Rev. B} {\bf 79}, 064412 (2009).

\bibitem{berezinskiikosterlitzthouless}
Berezinskii,~V. L. 
\newblock Destruction of long-range order in one-dimensional systems having a continuous symmetry group I. Classical systems. 
{\it Sov. Phys. JETP} {\bf 32}, 493 (1971);
Kosterlitz,~J.~M.~ and Thouless,~D.~J.
\newblock Ordering, metastability and phase transitions in two-dimensional systems. {\it J. Phys. C} {\bf 6}, 1181 (1973).

\bibitem{horowitz}
Horowitz,~B.,~  Bohr,~T.,~ Kosterlitz,~J.~M.~ and Schulz,~H.~J.
\newblock Commensurate-incommensurate transitions and a floating devil's staircase. {\it Phys. Rev. B} {\bf 28}, 6596 (1983).

\bibitem{wen}
Wen,~X. G. 
\newblock Chiral Luttinger liquid and the edge excitations in the fractional quantum Hall states. {\it Phys. Rev. B} {\bf 41}, 12838 (1990); 
{\it ibid.}, Gapless boundary excitations in the quantum Hall states and in the chiral spin states. {\it Phys. Rev. B} {\bf 43}, 11025 (1991); 
{\it ibid.}, Edge transport properties of the fractional quantum Hall states and weak-impurity scattering of a one-dimensional charge-density wave. {\it Phys. Rev. B} {\bf 44}, 5708 (1991).

\bibitem{lalsen}
Lal,~S.,~Rao,~S.~ and Sen,~D.  
\newblock Conductance through contact barriers of a finite-length quantum wire. {\it Phys. Rev. B} {\bf 65}, 195304 (2002).

\bibitem{polyakov}
Polyakov,~A.~M.  
\newblock Interaction of goldstone particles in two dimensions. Applications to ferromagnets and massive Yang-Mills fields. 
{\it Phys. Lett. B} {\bf 59}, 79 (1975).

\bibitem{haldane}
Haldane,~F.~D.~M.  
\newblock Continuum dynamics of the 1-D Heisenberg antiferromagnet: Identification with the $O(3)$ nonlinear sigma model. 
{\it Phys. Lett. A} {\bf 93}, 464 (1983);
{\it ibid.}, Nonlinear Field Theory of Large-Spin Heisenberg Antiferromagnets: Semiclassically Quantized Solitons of the One-Dimensional Easy-Axis N$\acute{\mathrm{e}}$el State. {\it Phys. Rev. Lett.} {\bf 50}, 1153 (1983);
Affleck,~I.~ and Haldane,~F.~D.~M. 
\newblock Critical theory of quantum spin chains. {\it Phys. Rev. B} {\bf 36}, 5291 (1987).

\bibitem{afflecknpb}
Affleck,~I.
\newblock Exact critical exponents for quantum spin chains, non-linear $\sigma$ models at $\theta =\pi$ and the quantum Hall effect. 
{\it Nucl. Phys. B} {\bf 265}, 409 (1986). 

\bibitem{senlal}
Sen,~D.~ and Lal,~S. 
\newblock One-dimensional fermions with incommensuration. {\it Phys. Rev. B} {\bf 61}, 9001 (2000); 
Tanaka,~A.~and~Hu,~X.
\newblock Effective field theory with a $\theta$-vacua structure for two-dimensional spin systems. 
{\it Phys. Rev. B} {\bf 74}, 140407 (R) (2006).

\bibitem{goldenfeld}
Goldenfeld,~N. 
\newblock {\it Lectures on Phase Transitions and the Renormalization Group}, (Addison-Wesley, 1992).

\end{thebibliography}
\end{document}